\documentclass[%
prd,nofootinbib,showpac,
superscriptaddress,
preprint
, amsmath,amssymb
]{revtex4-1}

\usepackage{afterpage}

\usepackage{mathptmx}
\usepackage{hyperref}
\hypersetup{
	colorlinks=false,
	pdfborder={0 0 0}
}
\def\Mr{\uppercase}

\usepackage{lineno}
\usepackage{float}
\usepackage{wasysym}
\usepackage{graphicx,wrapfig,tikz}
\usepackage{caption}
\usepackage{subcaption}
\usepackage[mathscr]{eucal}
\usepackage{epstopdf}
\usepackage{array}
\usepackage{multirow}
\usepackage{ulem}
\usepackage{array}

\def\vsm{\vskip0.1cm}
\def\titles#1{\title{\large\bf\noindent #1}}
\def\authors#1{\author{\begin{flushleft}{#1}\end{flushleft}}}
\def\authord#1#2{\indent\Mr{#1}\\
	\textit{\indent#2}\vsm}

\def\email#1{\bigskip\href{mailto:#1}{\textit{E-mail:}~{#1}}\\[3mm]}

\def\and{$\text{\tiny AND }$}

\usepackage[markup=underlined]{changes}

\begin{document}
\titles{Practical use of reactor anti-neutrinos for nuclear safeguard in Vietnam}
\authors{	\authord{S. Cao$^{1}$, T. V. Ngoc$^{1,2}$,  N. T. Hong Van$^3$, P. T. Quyen$^{1}$} 
	{
		\large{${}^{(1)}$ Institute For Interdisciplinary Research in Science and Education, Quy Nhon, Vietnam}\\[0.1cm] %
	\large{${}^{(2)}$ Graduate University of Science and Technology, Vietnam Academy of Science and Technology, Hanoi, Vietnam}\\[0.1cm]
	\large{${}^{(3)}$ Institute of Physics, Vietnam Academy of Science and Technology, Hanoi, Vietnam}\\ 
		\email{cvson@ifirse.icise.vn, tranngocapc06@ifirse.icise.vn}
		}
	}

	

\date{\today}	


\begin{abstract}
One of the most abundant man-made sources of low energy (few~MeVs) neutrinos, reactor neutrino, is not only useful for studying neutrino properties, but it is also used in practical applications. 
In this study, we investigate the practical use of reactor neutrino detectors for nuclear safeguard in Vietnam, specifically at the Dalat Nuclear Reactor, a future research facility, and presumably commercial reactors with 500~kW, 10~MW, and 1000~MW thermal powers, respectively. We compute the rate of observed inverted beta decay events, as well as the statistical significance of extracting isotope composition under the practical assumptions of detector mass, detection efficiency, and background level. We find that a 1-ton detector mass can allow us to detect the reactor's on-off transition state from a few hours to a few days, depending on the standoff distance and reactor thermal power. We investigate how background and energy resolution affect the precision of the extracted weapon-usable ${}^{239}\text{Pu}$ isotope. We conclude that in order to distinguish the 10\% variation of the ${}^{239}\text{Pu}$ in the 10~MW thermal power reactor, a 1-ton detector placed 50~m away must achieve 1\% background level. Increasing the statistics by using a 10x larger detector or placing it $\sqrt{10}$ times closer to the reactor alleviates the requirement of the background level to 10\%.
\end{abstract} 

	\maketitle

\section{Applied anti-Neutrino detection: a ``ghost" particle for peace}
In 1930, W. Pauli~\cite{Pauli:1930pc} proposed the existence of neutrino \textemdash a neutral lepton particle \textemdash to address the continuity of the beta decay spectrum, and it was experimentally detected for the first time in a reactor-based experiment twenty-six years later~\cite{Reines:1956rs}. The neutrino interacts only weakly in the Standard Model of elementary particles and is predicted to be massless because only left-handed neutrinos have been observed experimentally~\cite{Goldhaber:1957zz}. The fact that neutrinos are massive, as indicated by neutrino oscillations~\cite{RevModPhys.88.030501,RevModPhys.88.030502}, is experimental evidence that defies explanation by the Standard Model. Many efforts, summarized in Ref.~\cite{Zyla:2020zbs}, have been made over nearly a century to understand the nature of neutrinos. However, neutrinos remain one of the universe's most enigmatic particles. In fact, we still do not know whether neutrinos are Dirac or Majorana particles. We do not know how neutrinos can have an extraordinary smallness of mass, at least one million time smaller than the electron's mass. Despite some exciting hints~\cite{T2K:2019bcf}, there is insufficient statistical evidence to conclude that neutrino interactions violate the charge-conjugation and parity symmetry. The current state and future prospects of neutrino physics can be found, for example, in Ref. ~\cite{Zyla:2020zbs, Athar:2021xsd}.

Since neutrinos interact weakly with matter, they can pass through the earth, the sun, and the stars unaffected. On the one hand, it makes neutrino detection extremely difficult. Typically we require a large detector and a powerful neutrino source to make neutrinos visible. But on the other hand, it ushers in a new era of applied neutrino physics with reactor non-intrusive monitoring. Neutrino emitted by a reactor is electron anti-neutrino $\overline{\nu}_e$; we will refer to it as \textit{reactor neutrino} from now on. The reactor neutrinos are commonly detected using the inverse beta decay (IBD), which is discussed in the Section~\ref{sec:flux_detection}. With recent advances in detection technology and our progress in understanding reactor neutrino production, we can remotely monitor reactor thermal power and measure isotope composition using neutrinos as a working tool~\cite{Akindele:2021sbh}. Plutonium ${}^{239}\text{Pu}$, one of the reactor's four main isotopes, can be used to make nuclear weapons. As a matter of fact, continuous monitoring of the plutonium content in the reactor is critical for nuclear safety and verification regulations.

Since the 1980s, there has been discussion~\cite{Borovoi:1978} about the practical application of measuring reactor neutrino production to monitor the reactor itself. Further research in the 199s and the early 200s~\cite{Klimov:1994,Declais:1994su,Bernstein-2002,Bernstein-2008} confirmed the direct relationship between the reactor thermal power and the number of reactor neutrinos.  To reduce background caused by the cosmic rays, the detectors, which were typically cubic meters in size, were placed a few tens of meters underground. One of the earliest efforts to realize the ground-level reactor neutrino-based monitor was developed at Japan's fast reactor JOYO~\cite{Furuta:2011iu}. However, the presence of a large background prevents this experiment from definitively reporting the reactor's on-off transition.  A significant progress and now becoming a benchmark~\cite{Carr:2018tak} for the neutrino-based reactor monitoring come from the PROSPECT experiment~\cite{PROSPECT:2018dtt}. PROSPECT can achieve signal-to-background (S/B) ratio of 1:1 with an overburden of approximately one meter-water-equivalent (mwe). There are a large number of efforts around the world to realize this emerging neutrino-based technology, which are summarized in Table~\ref{tab:reactornuexp}. The historical development of this field can be found in Ref.~\cite{Bernstein2020,Akindele:2021sbh}. In terms of detection technology, liquid scintillator (LS) and plastic scintillator (PS) are commonly used to detect the prompt $e^+$ signals produced by the IBD in conjunction with neutron. The water Cherenkov (WC) detectors with Gd-doped are occasionally chosen. There are several advantages to using a liquid scintillator, including high light yield, low energy threshold, homogeneity, purification, large-detector scalability, and cost-effectiveness. However, because they are highly flammable, poisonous, and chemically dissolved, LS are difficult to handle and operate. When a mobile neutrino detector for nuclear safeguards is considered, the situation becomes even more difficult. PS-based detectors have emerged as a viable option for safer above-ground deployment. Furthermore, the segmentation approach allows us to measure both the temporal and spatial correlations of the IBD-induced signal, thus enhance the S/B performance. The Gadolinium (Gd) or Lithium (Li) are normally deployed to increase the neutron-captured efficiency. Some detectors are equipped with the pulse-shape discrimination (PSD) capability to distinguish the neutron-induced signals from the gamma background. Some have the ``movable" capability, which is essential for the above-ground safeguard. 
  \begin{center}
\begin{tabular}{| m{3cm}| m{1.5cm} | m{1.cm} | m{1.5cm} | m{6cm} |}
\hline
Experiment & Country  & Com. Year & Dist. [m]  & Detection Specification\\
\hline
SONGS1~\cite{bowden2007experimental} & US & 2003  &20 & 0.6-ton Gd-LS; 25~mwe overburden  \\
SONGS~\cite{reyna2012compact} & US & 2009  &50 & PS + $^6{\text{Li}}$; ground-level; movable   \\
\hline
At JOYO~\cite{Furuta:2011iu} & Japan & 2007 & 24.3 & 0.76-ton Gd-LS; ground-level \\
\hline
Cormorad~\cite{battaglieri2010anti} & Italy & 2009 & 25 & PS w/ Gd-coated sheet\\
\hline
PANDA~\cite{Kuroda:2012dw} & Japan & 2011  & 36 & PS w/ Gd-coated sheet; ground-level; movable \\
\hline
Nucifer~\cite{boireau2016online} & France & 2012  & 7 & 0.8-tons Gd-LS; 12~mwe overburden; PSD capability\\
\hline
Liverpool~\cite{carroll2018monitoring} & UK & 2014  &60 & 1-ton PS + Gd-coated sheet; movable   \\
VIDARR~\cite{coleman2019vidarr} & UK &   &Any & 1.5-ton PS + Gd-coated sheet; movable    \\
\hline
Nulat~\cite{dorrill2019nulat} & US & 2015  & 3-6 & 1-ton PS + $^6{\text{Li}}$ \\
\hline
WATCHMAN~\cite{askins2015physics} & US,UK & 2015  & 25000 & 1-ktons Gd-doped WC\\
\hline
Solid~\cite{abreu2018performance} (\cite{abreu2021solid}) & Belgium  & 2015 (2018)  & 6-9 & 288kg (1.6-tons) PS cube LiF-ZnS; PSD capability \\
\hline
DANSS~\cite{alekseev2016danss} & Russia & 2016  & 10.7-12.7 & 1-ton, PS-Gd, Movable, 50~mwe overburden\\
\hline
NEOS~\cite{ko2017sterile} & South Korea & 2017  & 23.7 & 1-ton Gd-LS; 25~mwe overburden; PSD capability \\
\hline
MiniCHANDLER & US & 2017  & Any & 80~kg PS + $^6{\text{Li}}$; ground-level \\
CHANDLER~\cite{Haghighat:2018mve} & US &   & Any & 1-ton PS + $^6{\text{Li}}$; ground-level \\
\hline
Angra~\cite{lima2019neutrinos} & Brazil & 2017  & Any & 1.34-ton GdCl3 WC; ground-level \\
\hline
STEREO~\cite{almazan2018sterile} & France & 2018  & 10.3 & 1.8-ton, segmented Gd-LS; 10~mwe overburden; PSD capability \\
\hline

PROSPECT~\cite{PROSPECT:2018dtt} & US  & 2018  & 7 & 4-tons segmented Li-LS; $\leq$1~mwe overburden; PSD \\
\hline
Neutrino-4~\cite{serebrov2019first} & Russia  & 2019  & 6-12 & 1.42m3, segmented Gd-LS; movable; $\sim$10~mwe overburden \\
\hline
\end{tabular}
\vspace*{-2mm}
\captionof{table}{Reactor-based projects developed in the 20th century targeting the reactor monitoring. Some projects have developed for fundamental physics for searching the sterile neutrino at very short-baseline neutrino experiment }\label{tab:reactornuexp}
\end{center}
To sum it up, the technology for the neutrino-based reactor monitors is ready for the commercialization and deployment~\cite{Akindele:2021sbh}.  Further research is needed to refine the technology and optimize it for cost, detection efficiency, and portability. The reactor-based neutrino experiments have played important role in exploring the fundamentals of neutrino particle, including but not limited to the discovery of neutrino~\cite{Reines:1956rs} and precision of mixing angle $\theta_{13}$~\cite{Zyla:2020zbs}. Still the medium-baseline experiment like JUNO~\cite{JUNO:2015zny} can help us to determine the neutrino mass ordering, which is one the unanswered questions of the neutrino physics. The short baseline experiment was designed to not only provide a precise understanding of the reactor neutrino flux, but also to search for the sterile neutrino, as seen in Ref.~\cite{Berryman_2021}. There is a strong synergy between the applied antineutrino physics for non-destructive monitoring and the basic science research with the reactor neutrinos.

\section{{Reactor neutrino flux and detection methodology}}\label{sec:flux_detection}
Understanding the neutrino produced by the reactor is critical for both basic science and practical applications of the reactor neutrinos. It is recalled that reactor neutrinos are produced from the $\sim 10,000$ beta decay branches of the $\sim 800$ fission fragments, which are generalized with ${}^A_{Z}X\rightarrow {}^A_{Z+1}X +e^{-}+\overline{\nu}_e$. Neutrinos are produced along with the associated beta particles. For this reason, there are two approaches for calculating the neutrino production from the reactor. The first approach is to sum up all beta-decay branches from all fission fragments. An updated summation model, called \textit{ab initio}, can be found in Ref.~\cite{estienne2019updated}. The second approach is to extract the the reactor neutrino spectrum from the $\beta$-spectral measurements. For the most of commercial reactors, there are four main isotopes ${}^{235}\text{U}, {}^{238}\text{U}, {}^{239}\text{Pu}$, and ${}^{241}\text{Pu}$, which contribute to approximately $99.9\%$ of the total reactor neutrino flux. The lifetime, natural abundance, number of reactor neutrinos per fission, and the released thermal energy per fission for individual isotopes are shown in Table~\ref{tab:isotope}.
\begin{center}
\begin{tabular}{|c| c| c| c|c|}
\hline
Isotope & ${}^{235}\text{U}$  & ${}^{238}\text{U}$ & ${}^{239}\text{Pu}$  & ${}^{241}\text{Pu}$\\
\hline
Lifetime[year] & $7\times 10^8$  & $1.6\times 10^5$ & $2.4\times 10^4$ & 14  \\
\hline
Natural abundance [$\%$] & 0.72 & 0.0 & 0.0 & 0.0\\
\hline
$N_{\bar{\nu}_e}$ per fission & $1.92\pm 0.04$ & $2.38\pm 0.05$  & $1.45\pm 0.03$ & $1.83\pm 0.03$\\
\hline
$E_{\text{released}}$  per fission [MeV] & $201.7\pm0.6$ & $205.0\pm0.9$ & $210.0\pm0.9$ & $212.4\pm1.0$\\
\hline
\end{tabular}
\vspace*{-2mm}
\captionof{table}{Four main isotopes in the reactor are presented with the lifetime, natural abundance, neutrino yield, and released thermal energy. Values are based on Ref.~\cite{patrick_huber_2022_6683772,nufluxHuber:2004}}\label{tab:isotope}
\end{center}
The $\beta$-spectral measurements for these isotopes have been carried out \cite{nuflux:1981,nuflux:1985,nuflux:1982,nuflux:1989,nufluxHaag2014} and there are two neutrino flux models~\cite{Mueller:2011,Huber:2011}~\cite{Hayen:2019eop} based on these measurements. The comparison among these flux models can be found in Ref.~\cite{Berryman_2021}. The uncertainty on the integrated flux of ${}^{235}\text{U}$, ${}^{239}\text{Pu}$, and ${}^{241}\text{Pu}$ is about 2-3$\%$ while the uncertainty on the integrated flux of ${}^{238}\text{U}$ is about 11$\%$. At given time $t$, when the fission fraction $f_i$ of the $i^{\text{th}}$ isotopes and their corresponding neutrino spectrum $S_i(E_{\overline{\nu}_e})$ is known, the total reactor neutrino flux $\Phi(E_{\overline{\nu}_e},t)$ can be summed up and normalized by the thermal power of the reactor $W_{\text{th.}}$ and the released thermal energy $e_i$ by the isotope: 

\begin{align}\label{eq:flux}
\Phi(E_{\overline{\nu}_e},t) = \frac{W_{\text{th.}}(t)}{\sum_if_i(t)e_i}\times \sum_if_i(t)\times S_i(E_{\overline{\nu}_e},t).
\end{align}

In average, as shown in Table~\ref{tab:isotope}, every fission releases about 200~MeV of energy and emits about 6 electron anti-neutrinos along the beta decay chain. For a reactor
with 1~$\text{GW}_{\text{th.}}$ (equivalent to $6.242 \times 10^{27}$ eV/s), the corresponding number of fission occur per second are computed to be $3 \times 10^{19} (\text{fissions})$. Thus, the total number of anti-neutrinos emitted by a 1~$\text{GW}_{\text{th.}}$ reactor are $\approx 2 \times 10^{20}$ particles per second. If the detector stands off at 50~m, it will receive a reactor neutrino flux of  $6\times 10^{11}\ \text{s}^{-1}\text{cm}^{-2}$. 

 \begin{figure}[H]
 	\centering
 \includegraphics[width=0.9\textwidth]{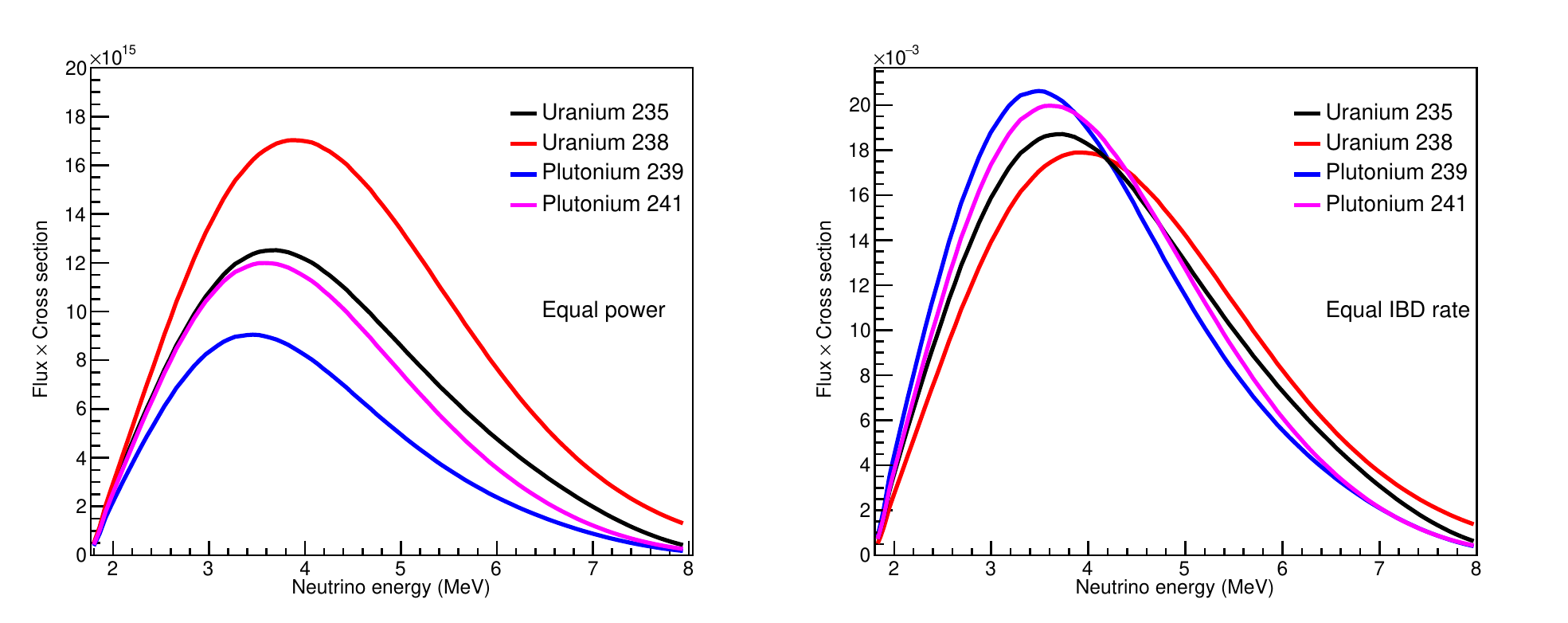}
\caption{Contribution to the reactor neutrino flux  from four main isotopes in term of the product of flux and cross-section as a function of neutrino energy. The left plot is with equal power and the right is with equal IBD rate. The ${}^{239}\text{Pu}$ and ${}^{241}\text{Pu}$ spectra are softer than the ${}^{235}\text{U}$ and ${}^{238}\text{U}$}
\label{fig:nuflux}
 \end{figure}

\noindent Fig.~\ref{fig:nuflux} shows the individual contributions from the four main isotopes to the total reactor neutrino flux. The comparison is on the product of the flux and IBD cross section and is shown in two cases (i) equal power and (ii) equal IBD rate. It can be observed that the ${}^{239}\text{Pu}$-contributed neutrino spectra are softer than the ${}^{235,238}U$-contributed neutrino spectra. As a result, the overall reactor neutrino spectrum will become softer when the fraction of ${}^{239}\text{Pu}$ increases. This is a key point for using reactor neutrino as an isotope spectroscopy, which is discussed in detail in Section~\ref{sec:resultpluto}. 

The observable rate of reactor neutrino as a function of energy $E_{\bar{\nu}_e}$ and time t, $N_{\bar{\nu}_e \ det.}(E_{\bar{\nu}_e},t)$ as shown in Eq.~\ref{eq:nuevent}, is proportional to the reactor neutrino flux $\Phi(E_{\bar{\nu}_e},t)$ defined in Eq.~\ref{eq:flux}, detector standoff $ \frac{1}{4\pi L^2}$, the cross section $\sigma (E_{\bar{\nu}_e})$, the number of targets of the detector $N_{\text{target}}$, the neutrino detection efficiency $\epsilon(E_{\bar{\nu}_e})$, and the survival probability of reactor neutrinos $P_{(\overline{\nu}_e\rightarrow\overline{\nu}_e)}(E_{\bar{\nu}_e}, L,\vec{o})$ where $\vec{o}$ presents the mixing parameters of the neutrino oscillation.
\begin{align}\label{eq:nuevent}
    N_{\bar{\nu}_e \ det.}(E_{\bar{\nu}_e},t) = \frac{\Phi(E_{\bar{\nu}_e},t)}{4\pi L^2} \times \sigma (E_{\bar{\nu}_e}) \times N_{\text{target}} \times \epsilon(E_{\bar{\nu}_e})\times   P_{(\overline{\nu}_e\rightarrow\overline{\nu}_e)}(E_{\bar{\nu}_e}, L,\vec{o}).
\end{align}
The IBD cross section, which about $5\times 10^{-43} \text{cm}^{-2}$ at 2~MeV neutrino energy, has been computed in Ref.~\cite{Vogel1999} with less than 1$\%$ uncertainty. The neutrino-based reactor monitoring can be classified into two types basing on the reactor-to-detector distance.
For near-field ($L\sim 10s$ m)  monitoring and if there is nonexistence of sterile neutrinos at eV scale, the oscillation probability can be approximated to be $P_{(\overline{\nu}_e\rightarrow\overline{\nu}_e)} \approx 1.0$. The effect of oscillation phenomenon should be taken into account with the far-field  ($L\sim 10s$ km)  monitoring. The IBD detection efficiency with LS-based detectors~\cite{PROSPECT:2018dtt,abreu2021solid} is slightly more than 40$\%$. For the segmented PS-based reactor mornitoring~\cite{Haghighat:2018mve}, the detector IBD efficiency reaches to approximately 60$\%$. In short, a 1-ton detector, which stand off 50~m from a 1~$\text{GW}_{\text{th.}}$ reactor, can detect few 10s of the IBD interactions per hour.

\begin{figure}[H]
 	\centering
 \includegraphics[width=0.7\textwidth]{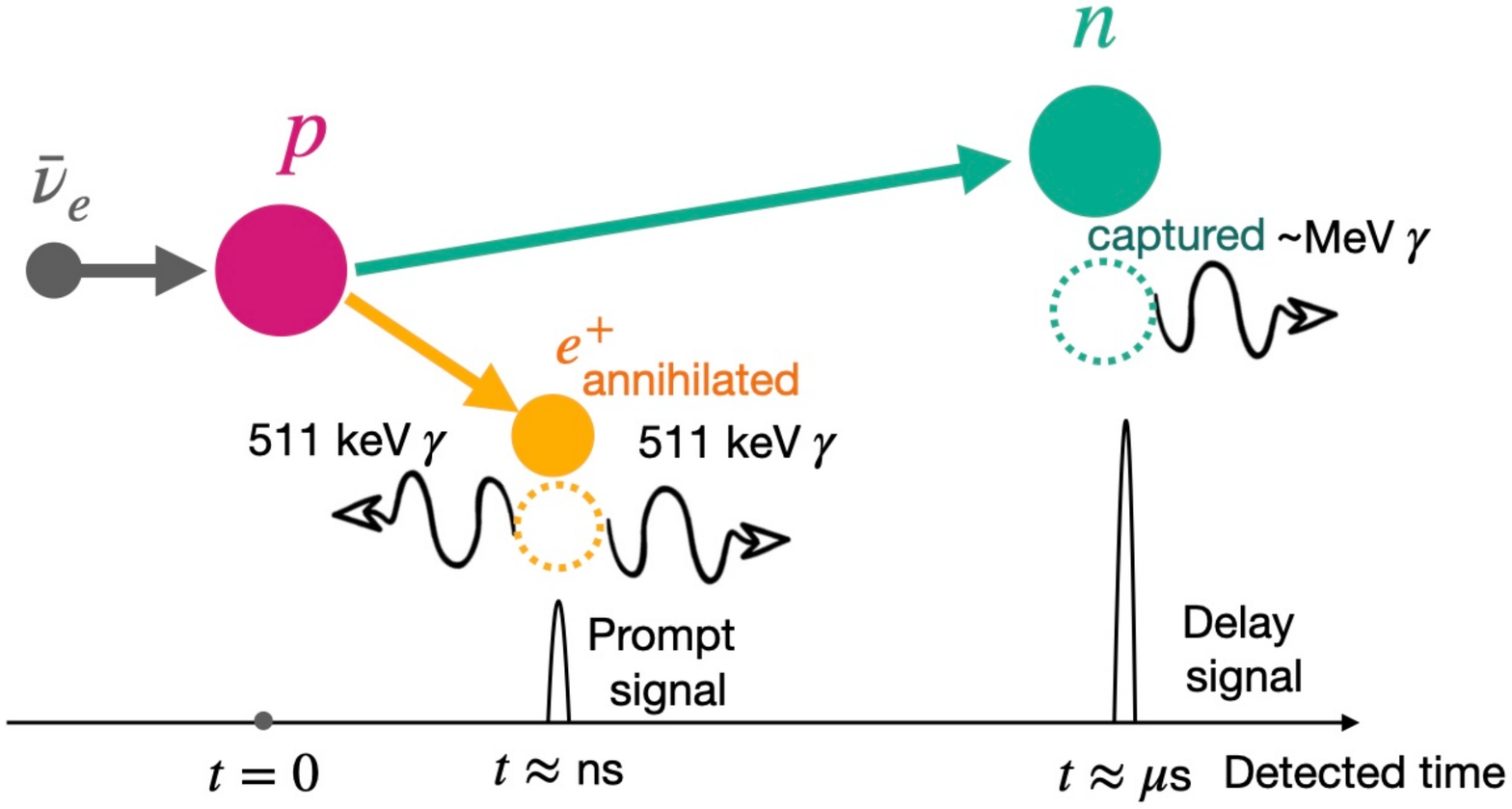}
\caption{Principle of the reactor neutrino detection via the inverse beta decay. A prompt signal comes from $e^+$ annihilation and a delayed signal is from the neutron capture. The gamma energy due to the neutron capture can be around 2.2~MeV in hydrogen or 8.0~MeV (4.8~MeV) with Gadolium (Lithium) respectively. Coincidence of the prompt signal and delay signal in a pre-defined timing window is a signal of the IBD interaction. }
\label{fig:ibddetect}
 \end{figure}

Although other methods, such as charged-current and neutral-current interactions with heavy waters, neutrino-electron elastic scatterings, and neutrino-nucleus coherent scattering, can be used to detect reactor neutrinos, the most common method is the aforementioned IBD, which produces simultaneously positron and neutron in the final state:
        \begin{align}
        \overline{\nu}_e + p \rightarrow e^{+} + n, \ (\text{when} \ E_{\overline{\nu}_e}>1.806\ \text{MeV}).
       \end{align}
The spatial and temporal coincidence of the highly detectable positron and neutron provides us with a distinguishing feature that allows us to effectively identify the reactor neutrinos.The threshold of this reaction is 1.806 MeV due to the difference in the ($e^{+}, n$) mass and ($\overline{\nu}_e, p$) mass. The spectrum of the reactor neutrinos can be deduced from the measured beta spectrum using the relation $E_\nu = E_{e^{+}} + 0.784$ MeV. Positron induced from the IBD quickly deposits its energy and annihilates with an electron of the medium to create two gamma rays with typical 511~keV prompt signals. Neutron can be captured by free proton or Hydrogen (H), gadolinium (Gd), or Lithium (Li) depending on the detector technology leaving a delay gamma cascade. The general principle for detecting the reactor neutrinos via the IBD reaction is illustrated in Fig.~\ref{fig:ibddetect}.

\noindent Most of the reactor developed for the safeguard purpose must be sensitive to see both the prompt signal and delay signal to identify the IBD interaction. For a particular detector, the light collection of the positron prompt signal determines the reconstructed energy resolution of the reactor neutrinos. This essentially depends on the scintillation light yield, photosensor coverage, and quantum efficiency of the photosensor. The LS-based detector can reach 1000 p.e. per 1~MeV deposition, resulting in  3-4$\%$ energy resolution at 1~MeV. The PROSPECT experiment~\cite{PROSPECT:2018dtt} has already achieved 4.5$\%$ energy resolution while JUNO~\cite{JUNO:2015zny}, a medium-baseline experiment for the basic science, can achieve 3$\%$ energy resolution. For the PS-based detector, the miniCHANDLER~\cite{Haghighat:2018mve} have achieved 10$\%$ energy resolution. Signature of the neutron capture is important to discriminate the reactor neutrino events from the backgrounds. Therefore, the IBD-based detectors are normally made by proton target materials. In the detector medium, neutron can be captured by a free proton or Hydrogen and subsequently produces an averaged 180 $\mu s$ delayed signal of 2.2~MeV gamma ray:
$$n + p \rightarrow d + \gamma\ (2.2\ \text{MeV}) $$
To increase the neutron-captured efficiency and shorten the delayed time, Gd is customarily loaded into the detector. Neutron is captured by Gd to produce 8.0~MeV gamma cascade with averagely $20\ \mu s$ delayed from the prompt $e^+$ signal.  
\begin{align*}
     n + {}^{155}\text{Gd}({}^{157}Gd) \rightarrow {}^{156}\text{Gd}^*({}^{158}\text{Gd}^*) \rightarrow {}^{156}\text{Gd} ({}^{158}\text{Gd}) + \gamma s\ (8.0\ \text{MeV}) 
\end{align*}
In the case of ${}^6_3\text{Li}$, neutron-capture process emits a 4.8~MeV $\alpha$ particle. 
\begin{align*}
    n+{}^6_3\text{Li} \rightarrow {}^3_{1} \text{He} + \alpha\ (4.8\ \text{MeV})
\end{align*}

The $\alpha$ ionization produces a relatively slow scintillation light. A detector with the PSD capability allows us to distinguish this neutron-captured signal from the huge amount of swift gamma-induced signal. The neutron-captured efficiency can exceed more than 70$\%$~\cite{PROSPECT:2018dtt,abreu2021solid,Haghighat:2018mve}. The overall IBD detection efficiency can reach to approximately 60$\%$.

 As illustrated in Sec.~\ref{sec:expectedrate} and Sec.~\ref{sec:resultpluto}, not only is signal detection improved, but background suppression is critical for the neutrino-based reactor monitor. Due to detector size requirements and budget constraints, active shielding is limited. Passive shielding with multi-layers of neutron-absorbing materials is widely used. One recent benchmark~\cite{PROSPECT:2018dtt} established a signal-to-background ratio of 1:1. Advancing reactor neutrino detection and shielding technology is required to go beyond this benchmark.
\section{{Reactor in Vietnam and experimental simulation}} \label{sec:simulation_results}

Vietnam currently has only one small research nuclear reactor with a nominal power of 500~kW, the Dalat Nuclear Reactor (DNR)~\cite{dalat2014}, which has been operational for 60 years. DNR has been used for neutron scattering, neutron activation analysis, material testing, medicine application, and biological science research, education, and training~\cite{futureCNR:2020}. Since 2011, DNR has been fully operating with the Low Enriched Uranium (LEU) fuel of $19.75\% \ ^{235}\text{U}$ at 500~kW power. However, due to its low power and facility limitations, this research reactor will be unable to meet the increasing demands of research and application in the future. In the near future, another 10~MW nuclear reactor for research purposes (RNR)~\cite{futureCNR:2020} is planned to be built. This new RNR will be equipped with LEU VVR-KN fuel. Aside from traditional applications, RNR is being developed for advanced research and applications such as neutron transmutation doping of silicon and neutron scattering material science research. According to a government plan~\cite{vnnupower2022} approved in 2007, 8~$\text{GW}_{\text{e.}}$ (gigawatt electrical\footnote{$\text{GW}_{\text{e.}}$ refers to the electric power produced by a generator. Reactor neutrino flux is proportional to the reactor thermal power, but not electrical power.}) nuclear power should be available by 2025. By 2030, the power is expected to increase to 15~$\text{GW}_{\text{e.}}$ with a total of 14 nuclear reactors. However, the National Assembly of Vietnam postponed the plans in 2016 due to rising costs and safety concerns. Vietnam's nuclear program was recently revived and is being considered for recovery in 2022. In this work, we assume a commercial reactor (CNR) with an average thermal power of 1000~MW for study.

In this study, the General Long Baseline Experiment Simulator (GLoBES)~\cite{Huber:2002mx} is used to simulate the reactor neutrino source and the detector setup.  The inputs for GLoBES are provided including reactor and detector information as follow. For the reactor, thermal powers are 500~kW, 10~MW and 1000~MW for DNR, RNR, and CNR, respectively. The flux and cross section spectra are parameterized as in \cite{nufluxHuber:2004,Vogel1999}. The energy window is selected in a range from the threshold 1.8~MeV to maximum 8.0~MeV. The detector is made from plastic scintillator with total target mass of 1~ton. The signal efficiency is considered to be either $11.6\%$ or $60\%$, which are based on the previous achievements in Ref.~\cite{Kuroda:2012dw} and Ref.~\cite{Haghighat:2018mve} respectively. There is no background specified in the simulation for simplicity. However, background is taken into account when exploring the physical potentials of the monitor. The detector setup is summarized as in Table \ref{tab:setup}.
\begin{center}
\begin{tabular}{|c| c| c| c|}
\hline
Detector Mass & Operation Time & $\overline{\nu}_e$ energy window & Signal efficiency \\
\hline
1 ton & 1 year & 1.8~MeV - 8.0~MeV & 11.6$\%$ - 60$\%$ \\
\hline
\end{tabular}
\vspace*{-2mm}
\captionof{table}{Nominal experimental configuration for simulation of the neutrino-based reactor monitor presented in this study.}\label{tab:setup}
\end{center}
\section{Expected rate of reactor neutrinos and time to on-off transition detection}\label{sec:expectedrate}
Firstly, we investigate the number of observable events at various distances for each reactor.
  \begin{figure}[H]
   \minipage{0.485\textwidth}
 	\centering
 \includegraphics[width=8cm]{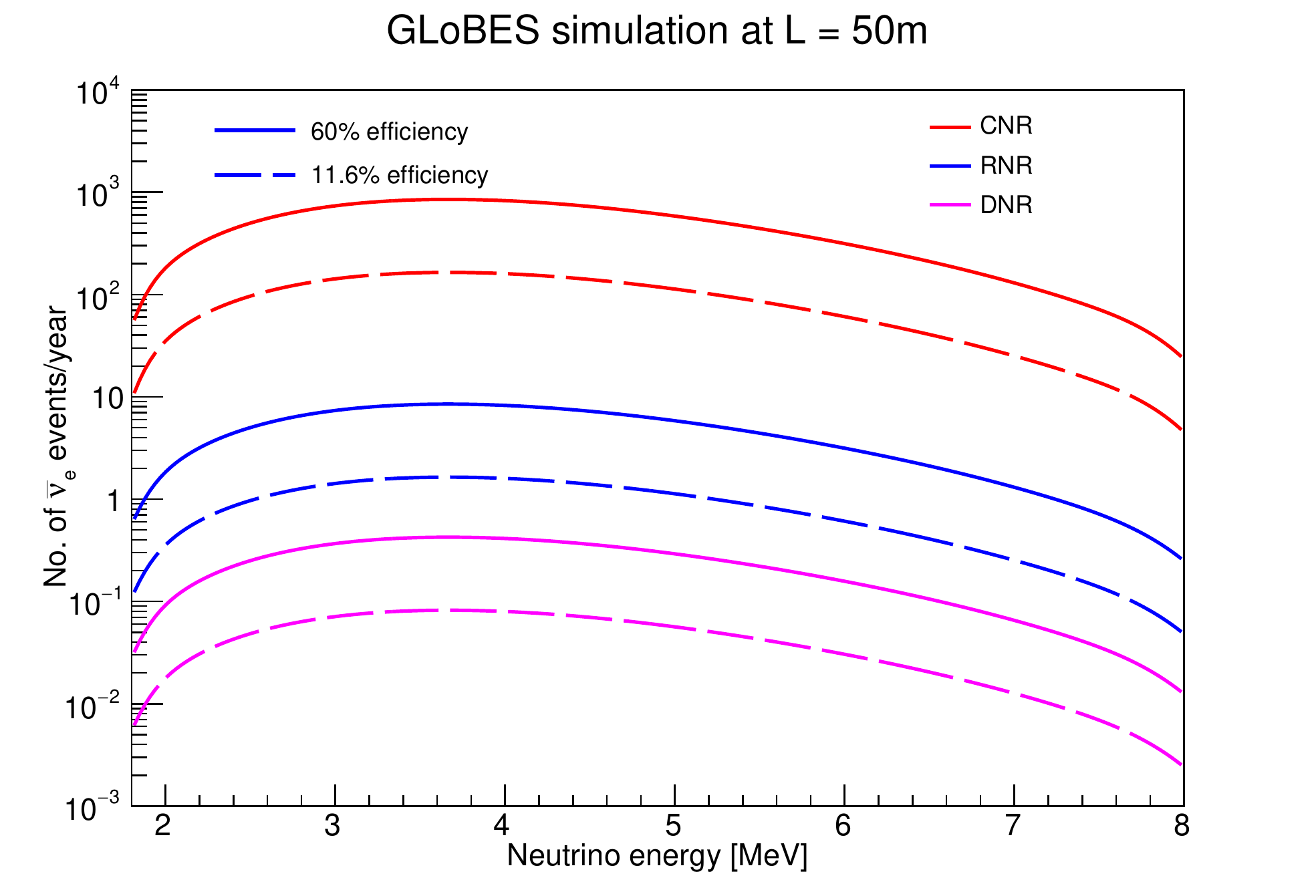}
 \caption*{}{(a)}
 \endminipage
 \hfill
\quad
 \minipage{0.485\textwidth}
 \centering
  	\includegraphics[width=8cm]{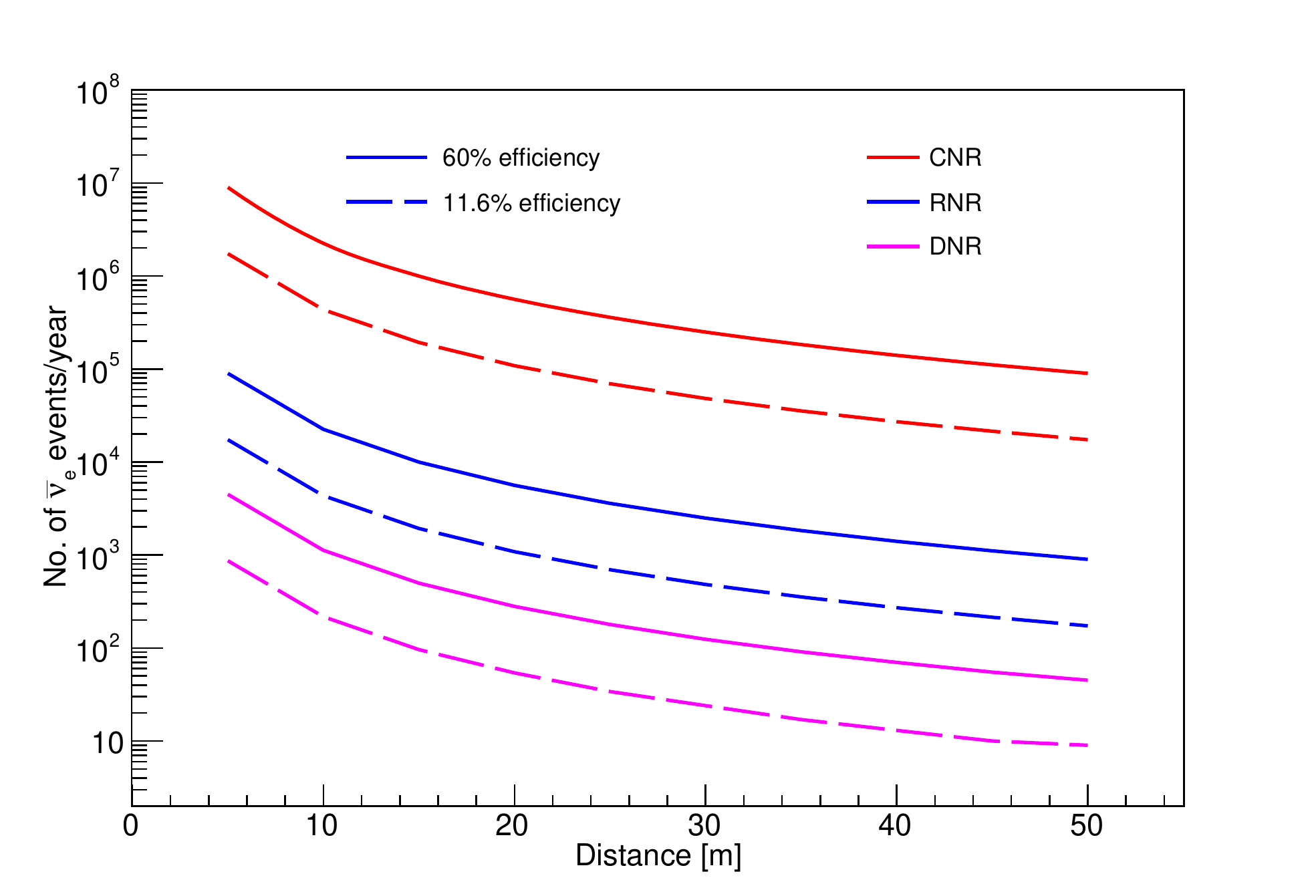} 
 	\caption*{}{(b)}
 \endminipage
  \caption{{\label{fig:eventrate}} The number of electron anti-neutrino events per year as a function of the (a) neutrino energy  and (b) distance from the reactor core for DNR (magenta line), future RNR (blue line) and CNR (red line). Solid lines and dashed are corresponding to $60\%$ and $11.6\%$ detection efficiencies.}
 \hfill
 \end{figure}
\noindent Fig.~\ref{fig:eventrate} shows the total number of observable anti-neutrino events in one year with respect to (a) the neutrino energy and (b) the distance from the reactor core.  The magenta, blue and red lines correspond to 500~kW DNR, 10~MW RNR and 1000~MW CNR respectively. The solid lines represent $60\%$ detection efficiency while the dashed lines are for $11.6\%$ efficiency. It can be seen from the Fig.~\ref{fig:eventrate} that the observable anti-neutrino spectra are maximal at around 4~MeV energy and inversely proportional to the distance square as described in Eq.~(\ref{eq:nuevent}). The Table~\ref{tab:eventrate} summarizes the expected annual IBD event rate at various baselines (5m, 10m, 20m, 50m). There are two numbers for the IBD event rate for each experimental configuration, which correspond to $11.6\%$ and $60\%$ detection efficiency, respectively. If the 1-ton detector is placed less than 20m from the DNR reactor core, the expected IBD event rate is in the hundreds to thousands per year. The expected IBD event rate for the RNR and CNR can reach $10^6$ collected per year.
\begin{center}
\begin{tabular}{|c| c| c| c| c|}
\hline
 Events/year & L=5~m &  L=10~m & L=20~m  &  L=50~m \\ 
 \hline  
 DNR & 	867 - 4487 & 217 - 1122 & 54 - 280 & 9	- 45	\\ 
  RNR & 17349 - 89734 & 4337 - 22433 & 1084 - 5608 & 173 - 897	\\ 
  CNR & $1.7 \times 10^6$  - $9.0 \times 10^6$ & 433714 - $2.2 \times 10^6$ & 108426 - 560826 & 17346 - 89720 \\
 \hline
\end{tabular} 
\captionof{table}{The expected IBD event rates from DNR, RNR and CNR at various baselines (5~m, 10~m, 20~m, 50~m) in one year. Two values in each cell are corresponding to the detection efficiencies of $11.6\%$ and $60\%$.}\label{tab:eventrate}
\end{center}

Assuming the signal-to-background (S/B) to be 1:1, achieved with PROSPECT~\cite{PROSPECT:2018dtt}, to detect the on-off transition of the reactor state at $5\sigma$ confidence-level (C.L.) statistical significance, the number of IBD events needed to collect is more than 30 events. The time to detection of the on-off transition state of the reactor can be calculated and shown in Table~\ref{tab:timedetection}.
\begin{center}
\begin{tabular}{|c| c| c| c| c|}
\hline
 Events/year & L=5~m &  L=10~m & L=20~m  &  L=50~m \\ 
 \hline  
 DNR & 	2.5 days & 10.1 days & 40 days & 251 days	\\ 
  RNR & 3hrs & 12 hrs & 2 days & 12.6 days	\\ 
  CNR & 1.8 mins. & 7.4 mins. & 29 mins. & 3 hrs \\
 \hline
\end{tabular} 
\captionof{table}{Time to detection of the on-off transition of the reactor state with 5$\sigma$ C.L for various experimental configuration. The detection efficiency is assumed to be $60\%$, and S/B = 1:1.}\label{tab:timedetection}
\end{center}
It has been observed that detecting the on-off transition of the reactor state with the DNR takes a few days to a month if the 1-ton detector is located less than 20~m. The RNR and CNR can detect this change in a matter of hours to days, even if the detector is placed 50~m away from the reactor core. For CNR, the time can be less than a half hour if the detector is placed within 20~m of the reactor core. The conclusion holds if we can achieve approximately 60$\%$ IBD detection efficiency and a S/B ratio of around 1:1 or better.
 
 \section{Potential to extract the plutonium content}\label{sec:resultpluto}
 To extract the isotope content, particularly the nuclear weapon-usable ${}^{239}\text{Pu}$, we rely on the reconstructed neutrino energy spectrum. Here, a 15$\%$ energy resolution is assumed without further mention throughout the work. The PS-based detector~\cite{Haghighat:2018mve} can achieve 10$\%$ energy resolution while the LS-based detectors~\cite{PROSPECT:2018dtt} can achieve 5$\%$ energy resolution. For this particular study, we assume to take a 1-year data with 1000~MW thermal power reactor for attaining statistics sufficiently for the analysis purpose. We implement a fit with five parameters, including a normalization factor $\alpha_0$, fractional ratio of Pu isotopes $\alpha_{1}$,  fractional ratio of ${}^{239}\text{Pu}$ in the Pu isotopes $\alpha_{2}$, fractional ratio of ${}^{235}\text{U}$ in the U isotopes $\alpha_{3}$, and a normalized background parameter $\alpha_{4}$. Assuming the nominal power-equal IBD event rates at the $i^{\text{th.}}$ energy bin with four main isotopes ${}^{235}\text{U}, {}^{238}\text{U}, {}^{239}\text{Pu}$, and ${}^{241}\text{Pu}$ are $N_{\text{ref.}}^{U-235,i}, N_{\text{ref.}}^{U-238,i}, N_{\text{ref.}}^{Pu-239,i}$, and $N_{\text{ref.}}^{Pu-241,i}$  respectively. Technically, we simulate these IBD rate with the 1:1:1:1 isotope composition. The total predicted number of the IBD-like events, including the signal and background, can be written as:
 \begin{align}
N_{\text{pred.}}^i=&\alpha_0 \times [N_{\text{ref.}}^{U-235,i}\times (1-\alpha_{1})\times \alpha_3+N_{\text{ref.}}^{U-238,i}\times (1-\alpha_{1})\times (1-\alpha_3) \nonumber \\ 
&+N_{\text{ref.}}^{Pu-239,i}\times \alpha_{1}\times \alpha_2 
+N_{\text{ref.}}^{Pu-241,i}\times \alpha_{1}\times (1-\alpha_2)]
+\alpha_{4}.
\end{align}
For a given isotope composition, for example ${}^{235}\text{U} \ : \ {}^{238}\text{U} \ : \ {}^{239}\text{Pu} \ : \ {}^{241}\text{Pu}$ = 4:1:4:1, we generate a fake data sample and retrieve the isotope composition by adopting the binned $\chi^2$ method to estimate the statistical significance. The $\chi^2$ function, a statistical measure of the difference between a predicted spectrum $N_{\text{pred.}}^{i}$ computed at a given set of parameters of interest and an observed spectrum $N_{\text{obs.}}^{i}$ (either real data or fake data), is defined as following
\begin{align*}
    \chi^2 =\sum_i  \frac{(N_{\text{pred.}}^{i}-N_{\text{obs.}}^{i})^2}{N_{\text{pred.}}^i . N_{\text{pred.}}^i}
\end{align*}
where $i$ is the energy bin index. The minimization of $\chi^2$ function allows us to find the best-fit set of parameters to describe the data and to attain the parameter intervals at the given statistical significance. 
 
We consider for two cases: (i) without the background and (ii) with the background. The former case study is to validate our framework to extract the isotope composition. 
 \begin{figure}[H]
   \minipage{0.485\textwidth}
 	\centering
  \includegraphics[width=7.5cm]{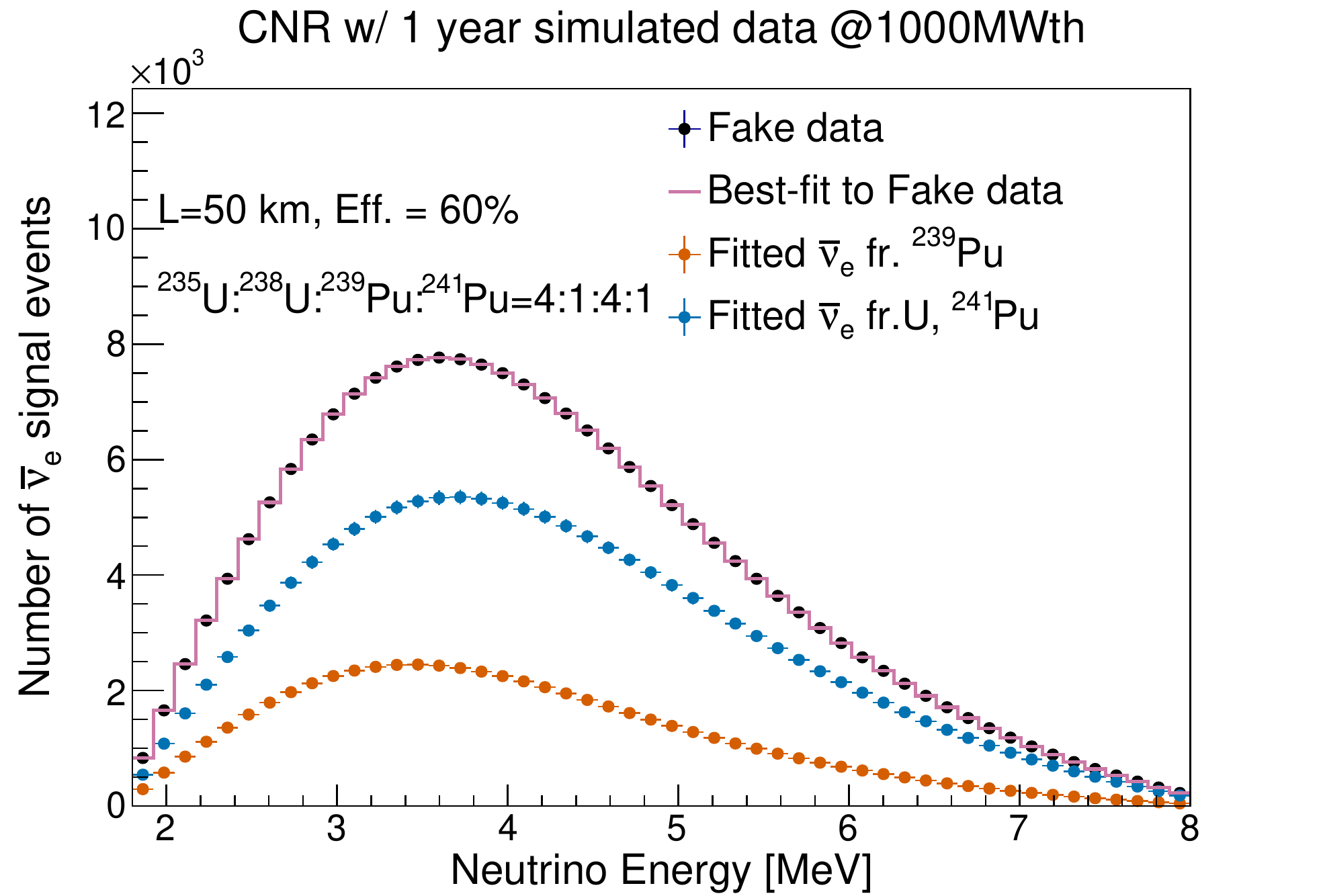}
 \endminipage
 \hfill
\quad
 \minipage{0.485\textwidth}
 \centering
  	\includegraphics[width=7.5cm]{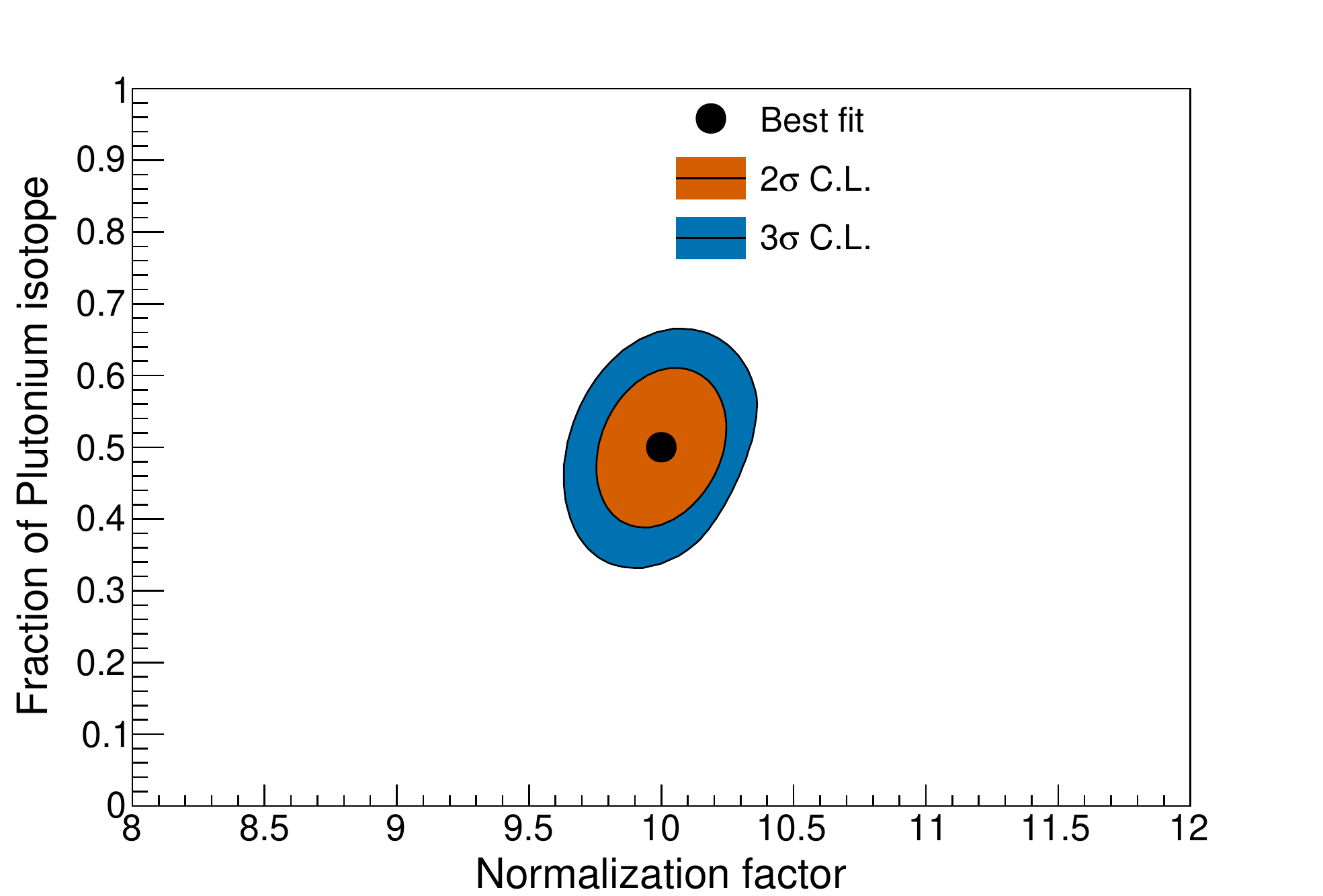} 
 \endminipage
  \caption{\label{fig:freebkgreactor50m60eff} On the left plot, a fake data in black points is generated  with an isotope composition assumed to be 4:1:4:1 and a $60\%$ IBD detection efficiency. Background is assumed to be free. The total best-fit spectrum shown the purple, which is summed from best-fit spectrum of ${}^{239} \text{Pu}$ in the orange and the other contributions in the blue, agree well with fake data. The right plot show 2$\sigma$ and 3$\sigma$ C.L. contour of $\alpha_0$ normalization factor and $\alpha_1$ fractional content of Plutonium isotopes. The best-fit value of $\alpha_1$ is 0.5, agree with an addition of 40\% ${}^{239} \text{Pu}$ and 10\% ${}^{241} \text{Pu}$ in the fake data.}
 \hfill
 \end{figure}
\noindent Fig.~\ref{fig:freebkgreactor50m60eff} present a case study with isotope composition ${}^{235}\text{U} \ : \ {}^{238}\text{U} \ : \ {}^{239}\text{Pu} \ : \ {}^{241}\text{Pu}$ = 4:1:4:1 without including the background. The total prediction at the best-fit parameters agrees with the fake data and the best-fit fractional composition of the Plutonium isotopes is 50$\%$, which agrees with an addition of 40$\%$ of ${}^{239}\text{Pu}$ and 10$\%$ of ${}^{241}\text{Pu}$. The 2$\sigma$ and 3$\sigma$ C.L. contours of two parameters, particularly $\alpha_0$ and $\alpha_1$ on the right plot of the Fig.~\ref{fig:freebkgreactor50m60eff},  are presented to give us the quantitative estimation of the fractional isotope composition.

\begin{figure}[H]
   \minipage{0.485\textwidth}
 	\centering
  \includegraphics[width=7.5cm]{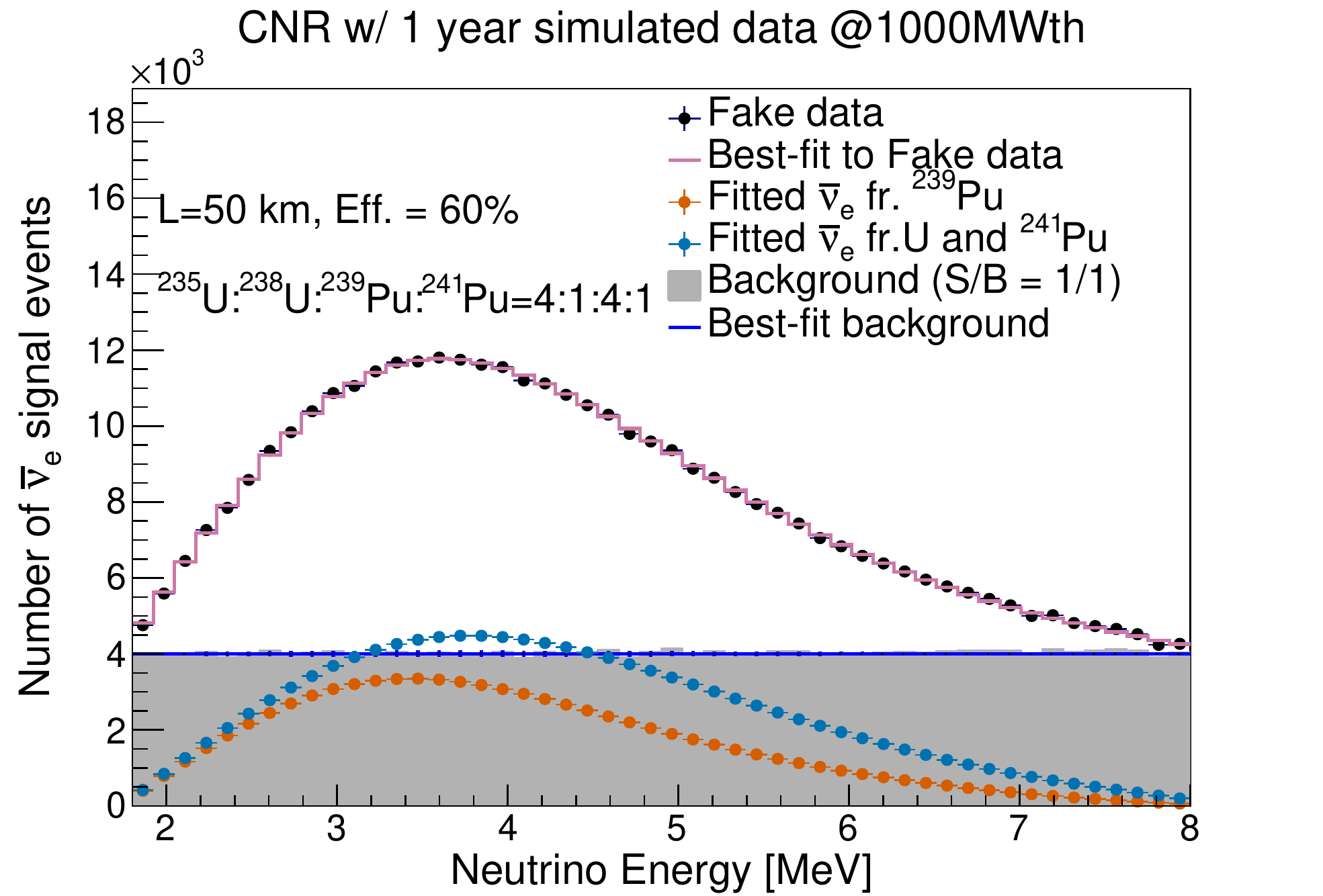}
 \endminipage
 \hfill
\quad
 \minipage{0.485\textwidth}
 \centering
  	\includegraphics[width=7.5cm]{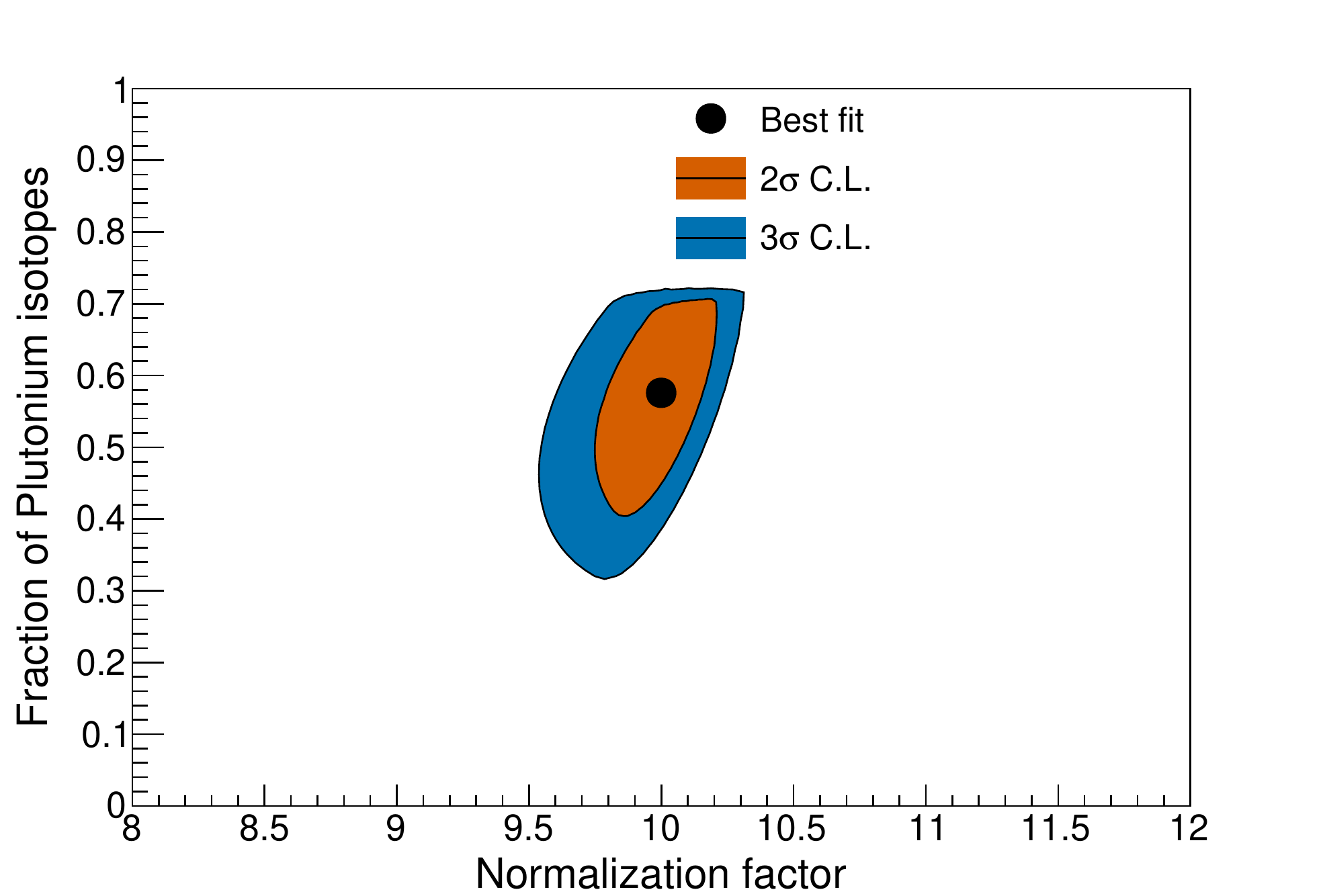} 
 \endminipage
  \caption{\label{fig:reactor50m60effsb11} The experimental setup is similar to the scenario presented in Fig.~\ref{fig:freebkgreactor50m60eff} except that background is included with a S/B=1:1. Background events, presented in grey shaded histogram, are fluctuated based on the Poisson statistics. The total best-fit spectrum shown the purple, which is summed from best-fit spectrum of ${}^{239} \text{Pu}$ in the orange, the contributions from other isotopes in the blue, and background in the blue solid-line, agree with the  data. The right plot shows 2$\sigma$ and 3$\sigma$ C.L. contours of $\alpha_0$ normalization factor and $\alpha_1$ fractional content of Plutonium isotopes. The best-fit value of $\alpha_1$ is slightly deviation from the expected value 0.5, but agrees statistically with an addition of 40\% ${}^{239} \text{Pu}$ and 10\% ${}^{241} \text{Pu}$ in the fake data. }
 \hfill
 \end{figure}

Apparently background-free detector is extremely challenging for the ground-level neutrino-based reactor monitor. Simulating the background is also difficult since it depends on the particular infrastructure of the reactor. A practical approach for background estimation is to measure with reactor-off state. Here we assume that the background is energy-independent for simplification. And we investigate for various scenarios of the S/B ratio. Fig.~\ref{fig:reactor50m60effsb11} shows a fit trial where we assume the S/B to be 1:1, achieved with PROSPECT experiment~\cite{PROSPECT:2018dtt},  and the Poisson-based statistical fluctuation of the background events is applied for individual bins of energy. 
 \noindent Overall the total best-fit spectrum, which is summed from the fitted IBD event rates from all isotopes and a constant background rate, describes relatively well the fake data. The effect of the statistically fluctuated background is observable in the 2-dimensional contour of the normalization factor $\alpha_0$ and the fraction of the Plutonium isotopes $\alpha_1$. The best-fit value of $\alpha_1$ is deviated from 0.5, which is expected as sum of 40\% of ${}^{239} \text{Pu}$ and 10\% ${}^{241} \text{Pu}$, but still agrees within 2$\sigma$ C.L. with the expectation. Another observable effect of the background is the enlargement and distortion of the 2-dimensional contour of the normalization factor and the fraction of Plutonium isotopes when compared to background-free case study in Fig.~\ref{fig:freebkgreactor50m60eff}. To investigate the background fluctuation statistically, we typically generate $10^{6}$ fake samples for each case of study by randomizing the background and fit independently. We then extract and accumulate the statistical distributions of the parameters of interest for the converged fit trials. The mean and root-mean-square (RMS) of the accumulated distribution are used to present the parameter of interest's central value and 1$\sigma$ uncertainty.

\begin{figure}[H]
   \minipage{0.485\textwidth}
 	\centering
 	  	\includegraphics[width=7.5cm]{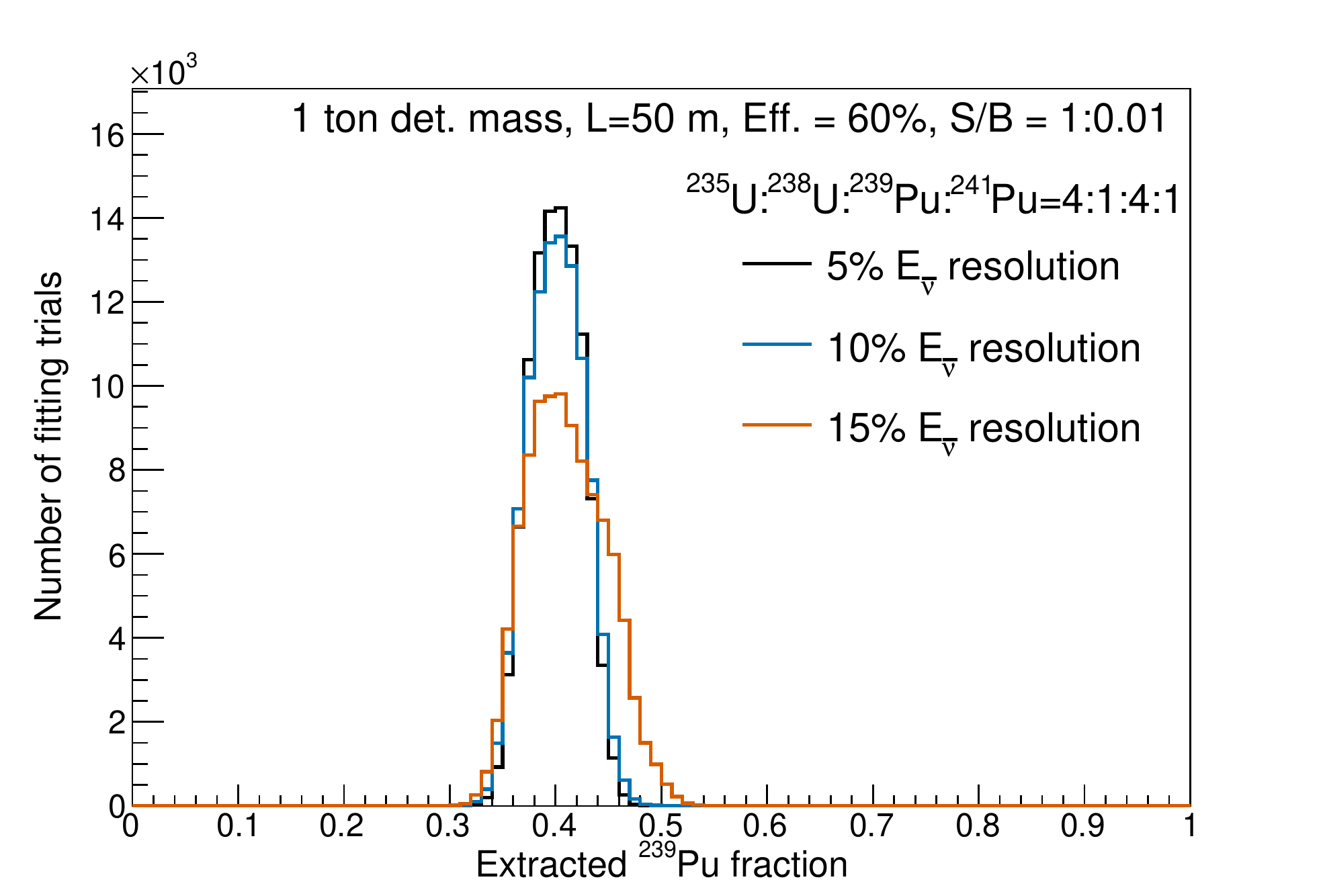} 
  
 \endminipage
 \hfill
\quad
 \minipage{0.485\textwidth}
 \centering
\includegraphics[width=7.5cm]{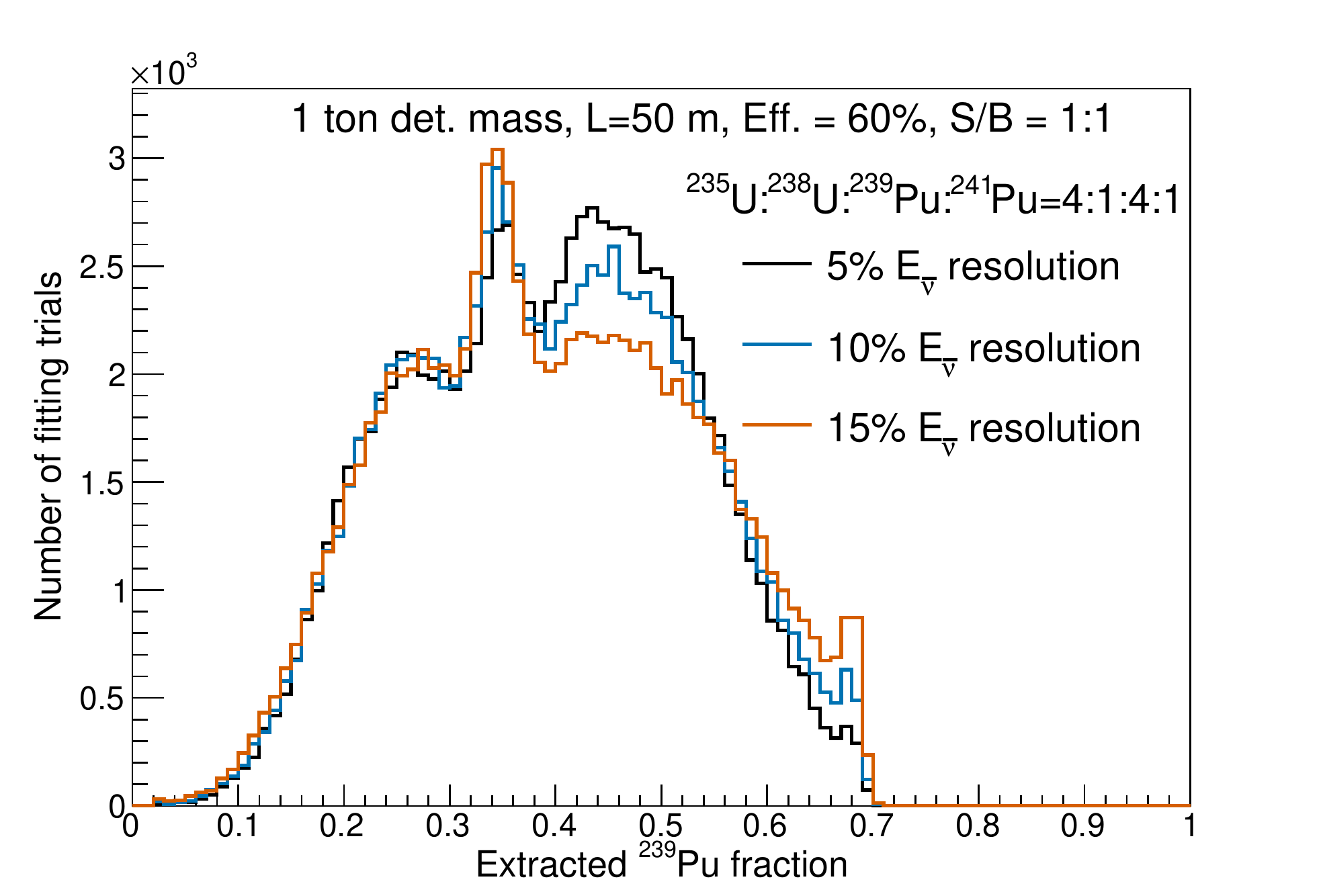}
 \endminipage
  \caption{\label{fig:fitbkgreseff}  Accumulated distributions of the extracted ${}^{239} \text{Pu}$ fraction with two S/B levels, S/B=1:0.01 (left) and S/B=1:1 (right), and three options of energy resolution (5\%, 10\%, and 15\%). The fake data is generated for a 1-ton detector placed at 50~m from the reactor core with the isotope composition ${}^{235}\text{U} \ : \ {}^{238}\text{U} \ : \ {}^{239}\text{Pu} \ : \ {}^{241}\text{Pu}$ = 4:1:4:1 with a $60\%$ IBD detection efficiency. For each configuration, $10^6$ fitting trials are performed on samples where Poisson-based randomized background events are included.}
 \hfill
 \end{figure}

In addition to the statistics of the observed IBD events, the precision of the extracted ${}^{239} \text{Pu}$ fractional content depends on the background level and reactor neutrino energy resolution. We investigate six experimental setups with two S/B levels (S/B=1:1 and S/B = 1:0.01) and three magnitudes of energy resolution (5\%, 10\%, and 15\%). The accumulated distributions of the extracted ${}^{239} \text{Pu}$ fraction are shown in Fig.~\ref{fig:fitbkgreseff} and precision of the extraction, which is attained from the RMSs of the distributions, is presented in Table~\ref{tab:eresefftwobkg}. 
 
 \begin{center}
\begin{tabular}{|c| c| c| c|}
\hline
S/B ratio & \multicolumn{3}{|c|}{Precision of the extracted ${}^{239}\text{Pu}$  } \\
\hline
& w/ 5$\%$ $E_{\nu}$ resol. &  w/ 10$\%$ $E_{\nu}$ resol. & w/ 15$\%$ $E_{\nu}$ resol.  \\ 
 \hline  
 S/B=1:1 & 	32.3$\%$ & 33.4$\%$ & 35.0$\%$ 	\\ 
 \hline
  S/B =1:0.01 & 6.0$\%$ & 6.5$\%$ & 9.2$\%$ 	\\ 
 \hline
\end{tabular} 
\captionof{table}{Precision of the extracted ${}^{239}\text{Pu}$ fractional content, attained from the RMS of the accumulated distribution in Fig.~\ref{fig:fitbkgreseff}, for two S/B levels, S/B=1:0.01 and S/B=1:1, and three options of energy resolution (5\%, 10\%, and 15\%).  }\label{tab:eresefftwobkg}
\end{center}
 
\noindent In case of the S/B=1:0.01, the precision of the extracted ${}^{239}\text{Pu}$ can range from 6.0\% to 9.2\%. The precision is less than 30$\%$ at the S/B = 1:1 level. The effect of energy resolution is visible, but the S/B level is the main driver for the precision of the extracted ${}^{239}\text{Pu}$ to reach few percent uncertainty, which is essential for tracking this weapon-usable isotope. 

\begin{figure}[H]
   \minipage{0.485\textwidth}
 	\centering
   	\includegraphics[width=7.5cm]{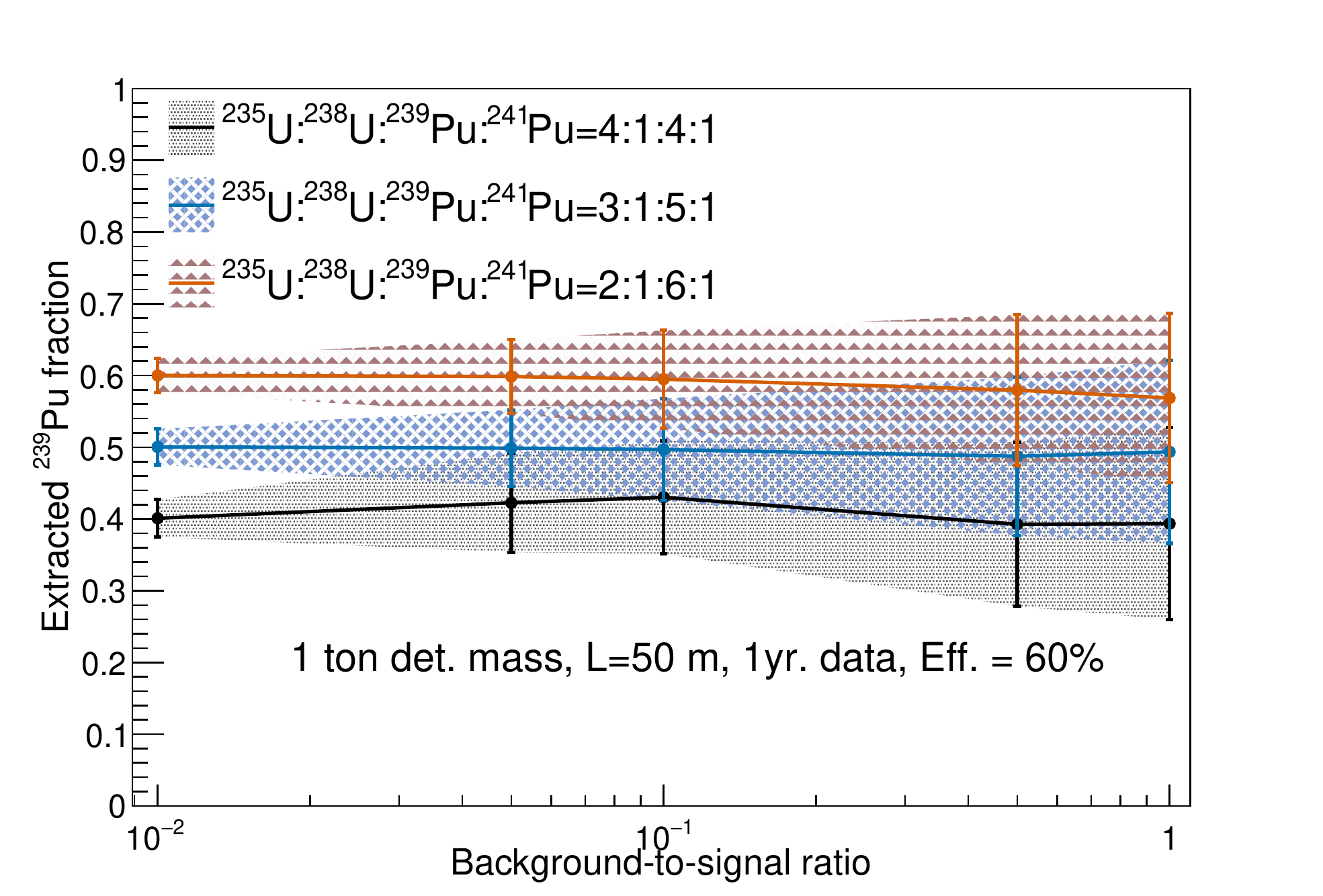} 
 \endminipage
 \hfill
\quad
 \minipage{0.485\textwidth}
 \centering
  	\includegraphics[width=7.5cm]{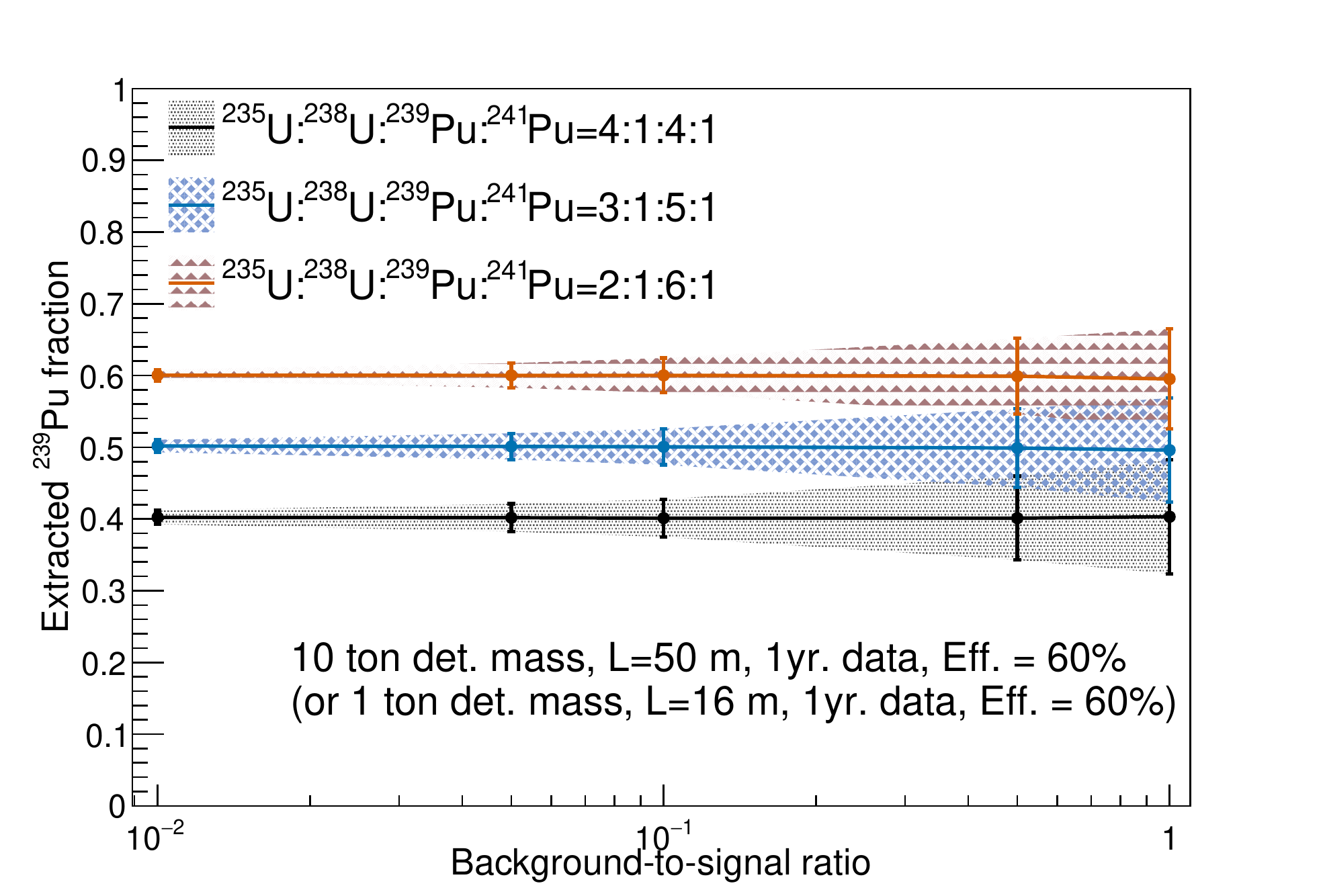} 
 \endminipage
  \caption{\label{fig:detscaleeff} The precision of extracted ${}^{239}\text{Pu}$ fraction as a function of the background-to-signal ratio ranged from 1\% to 100\% with three scenarios of the isotope composition: (i) ${}^{235}\text{U} \ : \ {}^{238}\text{U} \ : \ {}^{239}\text{Pu} \ : \ {}^{241}\text{Pu}$ = 4:1:4:1, (ii) ${}^{235}\text{U} \ : \ {}^{238}\text{U} \ : \ {}^{239}\text{Pu} \ : \ {}^{241}\text{Pu}$ = 3:1:5:1, and (iii) ${}^{235}\text{U} \ : \ {}^{238}\text{U} \ : \ {}^{239}\text{Pu} \ : \ {}^{241}\text{Pu}$ = 2:1:6:1. The left is with a 1-ton detector mass placed at 50~m standoff. The right is with a 10-ton detector at the same standoff, which is statistically equivalent to a 1-ton detector placed at 16~m standoff. }
 \hfill
 \end{figure}

 To investigate on the potential of tracking ${}^{239}\text{Pu}$ fraction dynamically, we study  three scenarios of isotope composition (i) ${}^{235}\text{U} \ : \ {}^{238}\text{U} \ : \ {}^{239}\text{Pu} \ : \ {}^{241}\text{Pu}$ = 4:1:4:1, (ii) ${}^{235}\text{U} \ : \ {}^{238}\text{U} \ : \ {}^{239}\text{Pu} \ : \ {}^{241}\text{Pu}$ = 3:1:5:1, and (iii) ${}^{235}\text{U} \ : \ {}^{238}\text{U} \ : \ {}^{239}\text{Pu} \ : \ {}^{241}\text{Pu}$ = 2:1:6:1, where the fractional content of ${}^{239}\text{Pu}$ varies from 40$\%$ to 60$\%$. In Fig.~\ref{fig:detscaleeff}, we show the precision of extracted ${}^{239}\text{Pu}$ fraction as a function of the background-to-signal ratio ranged from 1\% to 100\% with these three scenarios. We conclude that for a 1-ton detector placed 50~m away from the reactor core, the sensitivity to distinguish statistically the change in the three scenarios of the isotope composition is significant only when the background-to-signal is about 1\%. However, if we increase the statistics by either using a larger detector or placing the detector closer to the reactor core, for example with a 10-ton detector mass placed at the 50~m standoff or a 1-ton detector mass at the 16~m standoff, the neutrino-based reactor monitor enables to tell distinctively the variation of ${}^{239}\text{Pu}$ from 40\%, 50\%, and 60\% even with 10\% level of the background-to-signal ratio. 

\section{Summary}

The work is a preliminary investigation into using the neutrino-based reactor monitoring for safeguarding the reactors in Vietnam. The technique has been well-developed worldwide thank to the efforts from many reactor-based experiments. Our precise understanding of the reactor flux and advantage in the neutrino detection technology allows us to have a nearly real-time and non-destructive reactor monitoring with reactor neutrinos. We simulated and computed the expected reactor neutrino rate and time to detect the reactor's on-off state transition for the on-going Dalat research nuclear reator (DNR), the proposed 10~MW research facility (RNR), and the considered 1000~MW commercial reactors (CNR). We report that, based on the current performance of neutrino detection technology, the expected IBD event rate is in the hundreds to thousands per year and that detecting  the reactor's on-off state transition takes a few days to a month if the 1-ton detector is placed less than 20~m from the DNR reactor core. The annual IBD event rate for the future RNR and CNR facilities is expected to be $10^6$, allowing us to detect the on-off transition of the reactor state in a matter of hours to days, even if the detector is placed 50~m away from the reactor core. Another focus of this study is on the possibility of extracting the fractional content of the nuclear weapon-usable ${}^{239}\text{Pu}$ isotope. We show that with the 100\% background level, the 1-ton detector placed 50~m standoff can measure the ${}^{239}\text{Pu}$ fraction with a precision less than 30\% and that a 1$\%$ background level is needed to differentiate the 10\% variation of the ${}^{239}\text{Pu}$ fraction. The requirement for the background level can be loosened to 10\% level if we can increase the statistics by having 10x larger detector or place detector $\sqrt{10}$ times closer to the reactor core. 

Our work indicates that it is worthwhile to conduct research on the neutrino-based reactor monitor in Vietnam. While the technology as a whole is well-established around the world, more research into detection technology is needed to improve performance, lower costs, and make it more accessible. Our recent work with a novel type of silicon multiplier~\cite{Thanh:2021inp} demonstrates that this photosensor can be a good candidate for working with a plastic scintillator to build a monitor for this purpose, and our group plan to continue research in this area.
\section{Acknowledgement}

This research is funded by the Vietnam National Foundation for Science and Technology Development (NAFOSTED) under grant number 103.99-2018.45.
\bibliographystyle{unsrt}
\bibliography{reference}

\begin{thebibliography}{10}

\bibitem{Pauli:1930pc}
W.~Pauli.
\newblock {Dear radioactive ladies and gentlemen}.
\newblock {\em Phys. Today}, 31N9:27, 1978.

\bibitem{Reines:1956rs}
Frederick Reines and Clyde~L. Cowan.
\newblock {The neutrino}.
\newblock {\em Nature}, 178:446--449, 1956.

\bibitem{Goldhaber:1957zz}
M.~Goldhaber, L.~Grodzins, and A.~W. Sunyar.
\newblock {Evidence for Circular Polarization of Bremsstrahlung Produced by
  Beta Rays}.
\newblock {\em Phys. Rev.}, 106:826--828, 1957.

\bibitem{RevModPhys.88.030501}
Takaaki Kajita.
\newblock Nobel lecture: Discovery of atmospheric neutrino oscillations.
\newblock {\em Rev. Mod. Phys.}, 88:030501, Jul 2016.

\bibitem{RevModPhys.88.030502}
Arthur~B. McDonald.
\newblock Nobel lecture: The sudbury neutrino observatory: Observation of
  flavor change for solar neutrinos.
\newblock {\em Rev. Mod. Phys.}, 88:030502, Jul 2016.

\bibitem{Zyla:2020zbs}
P.~A. Zyla et~al.
\newblock {Review of Particle Physics}.
\newblock {\em PTEP}, 2020(8):083C01, 2020.

\bibitem{T2K:2019bcf}
K.~Abe et~al.
\newblock {Constraint on the matter\textendash{}antimatter symmetry-violating
  phase in neutrino oscillations}.
\newblock {\em Nature}, 580(7803):339--344, 2020.
\newblock [Erratum: Nature 583, E16 (2020)].

\bibitem{Athar:2021xsd}
M.~Sajjad Athar et~al.
\newblock {Status and Perspectives of Neutrino Physics}.
\newblock {\em arXiv hep-ph 2111.07586}, 11 2021.

\bibitem{Akindele:2021sbh}
Oluwatomi Akindele et~al.
\newblock {Nu Tools: Exploring Practical Roles for Neutrinos in Nuclear Energy
  and Security}.
\newblock {\em arXiv hep-ph 2112.12593}, 12 2021.

\bibitem{Borovoi:1978}
A.~A. Borovoi; L.~A. Mikaélyan.
\newblock Possibilities of the practical use of neutrinos.
\newblock {\em Atomic Energy}, 44:589--592, 1978.

\bibitem{Klimov:1994}
Y.~V. Klimov; V. I. Kopeikin; L. A.~Mikaélyan et~al.
\newblock Neutrino method remote measurement of reactor power and power output.
\newblock {\em Atomic Energy}, 76:123--127, 1994.

\bibitem{Declais:1994su}
Y.~Declais et~al.
\newblock {Search for neutrino oscillations at 15-meters, 40-meters, and
  95-meters from a nuclear power reactor at Bugey}.
\newblock {\em Nucl. Phys. B}, 434:503--534, 1995.

\bibitem{Bernstein-2002}
Adam Bernstein, Yi-fang Wang, Giorgio Gratta, and Todd West.
\newblock {Nuclear reactor safeguards and monitoring with anti-neutrino
  detectors}.
\newblock {\em J. Appl. Phys.}, 91:4672, 2002.

\bibitem{Bernstein-2008}
A.~Bernstein, N.~S. Bowden, A.~Misner, and T.~Palmer.
\newblock {Monitoring the Thermal Power of Nuclear Reactors with a Prototype
  Cubic Meter Antineutrino Detector}.
\newblock {\em J. Appl. Phys.}, 103:074905, 2008.

\bibitem{Furuta:2011iu}
H.~Furuta et~al.
\newblock {A Study of Reactor Neutrino Monitoring at Experimental Fast Reactor
  JOYO}.
\newblock {\em Nucl. Instrum. Meth. A}, 662:90--100, 2012.

\bibitem{Carr:2018tak}
Rachel Carr et~al.
\newblock {Neutrino-based tools for nuclear verification and diplomacy in North
  Korea}.
\newblock {\em arXiv physics.soc-ph 1811.04737}, 11 2018.

\bibitem{PROSPECT:2018dtt}
J.~Ashenfelter et~al.
\newblock {First search for short-baseline neutrino oscillations at HFIR with
  PROSPECT}.
\newblock {\em Phys. Rev. Lett.}, 121(25):251802, 2018.

\bibitem{Bernstein2020}
Adam Bernstein, Nathaniel Bowden, Bethany~L. Goldblum, Patrick Huber, Igor
  Jovanovic, and John Mattingly.
\newblock Colloquium: Neutrino detectors as tools for nuclear security.
\newblock {\em Rev. Mod. Phys.}, 92:011003, Mar 2020.

\bibitem{bowden2007experimental}
N.~S. Bowden et~al.
\newblock {Experimental results from an antineutrino detector for cooperative
  monitoring of nuclear reactors}.
\newblock {\em Nucl. Instrum. Meth. A}, 572:985--998, 2007.

\bibitem{reyna2012compact}
David Reyna.
\newblock A compact and portable antineutrino detector for reactor monitoring.
\newblock \url{https://www.osti.gov/biblio/1290207}, 2012.

\bibitem{battaglieri2010anti}
M.~Battaglieri, R.~De~Vita, G.~Firpo, P.~Neuhold, M.~Osipenko, D.~Piombo,
  G.~Ricco, M.~Ripani, and M.~Taiuti.
\newblock {An anti-neutrino detector to monitor nuclear reactor's power and
  fuel composition}.
\newblock {\em Nucl. Instrum. Meth. A}, 617:209--213, 2010.

\bibitem{Kuroda:2012dw}
Yasuhiro Kuroda et~al.
\newblock {A mobile antineutrino detector with plastic scintillators}.
\newblock {\em Nucl. Instrum. Meth. A}, 690:41--47, 2012.

\bibitem{boireau2016online}
G.~Boireau et~al.
\newblock {Online Monitoring of the Osiris Reactor with the Nucifer Neutrino
  Detector}.
\newblock {\em Phys. Rev. D}, 93(11):112006, 2016.

\bibitem{carroll2018monitoring}
J.~Carroll, J.~Coleman, M.~Lockwood, C.~Metelko, M.~Murdoch, Y.~Schnellbach,
  C.~Touramanis, R.~Mills, G.~Davies, and A.~Roberts.
\newblock {Monitoring Reactor Anti-Neutrinos Using a Plastic Scintillator
  Detector in a Mobile Laboratory}.
\newblock {\em arXiv physics.ins-det 1811.01006}, 11 2018.

\bibitem{coleman2019vidarr}
J.~Coleman, C.~Metelko, M.~Murdoch, Y.~Schnellbach, C.~Touramanis, R.~Mills,
  and D.~Mountford.
\newblock {VIDARR: Aboveground Reactor Monitoring}.
\newblock In {\em J. Phys. Conf. Ser.}, volume 1216, page 012007, 2019.

\bibitem{dorrill2019nulat}
Ryan Dorrill.
\newblock Nulat: A compact, segmented, mobile anti-neutrino detector.
\newblock In {\em J. Phys. Conf. Ser.}, volume 1216, page 012011. IOP
  Publishing, 2019.

\bibitem{askins2015physics}
M.~Askins et~al.
\newblock {The Physics and Nuclear Nonproliferation Goals of WATCHMAN: A WAter
  CHerenkov Monitor for ANtineutrinos}.
\newblock {\em arXiv physics.ins-det 1502.01132}, 2 2015.

\bibitem{abreu2018performance}
Y.~Abreu et~al.
\newblock {Performance of a full scale prototype detector at the BR2 reactor
  for the SoLid experiment}.
\newblock {\em JINST}, 13(05):P05005, 2018.

\bibitem{abreu2021solid}
Y.~Abreu et~al.
\newblock {SoLid: a short baseline reactor neutrino experiment}.
\newblock {\em JINST}, 16(02):P02025, 2021.

\bibitem{alekseev2016danss}
I.~Alekseev et~al.
\newblock {DANSS: Detector of the reactor AntiNeutrino based on Solid
  Scintillator}.
\newblock {\em JINST}, 11(11):P11011, 2016.

\bibitem{ko2017sterile}
Y.~J. Ko et~al.
\newblock {Sterile Neutrino Search at the NEOS Experiment}.
\newblock {\em Phys. Rev. Lett.}, 118(12):121802, 2017.

\bibitem{Haghighat:2018mve}
Alireza Haghighat, Patrick Huber, Shengchao Li, Jonathan~M. Link, Camillo
  Mariani, Jaewon Park, and Tulasi Subedi.
\newblock {Observation of Reactor Antineutrinos with a Rapidly-Deployable
  Surface-Level Detector}.
\newblock {\em Phys. Rev. Applied}, 13(3):034028, 2020.

\bibitem{lima2019neutrinos}
HP~Lima~Jr et~al.
\newblock Neutrinos angra experiment: commissioning and first operational
  measurements.
\newblock {\em JINST}, 14(06):P06010, 2019.

\bibitem{almazan2018sterile}
H.~Almaz\'an et~al.
\newblock {Sterile Neutrino Constraints from the STEREO Experiment with 66 Days
  of Reactor-On Data}.
\newblock {\em Phys. Rev. Lett.}, 121(16):161801, 2018.

\bibitem{serebrov2019first}
A.~P. Serebrov et~al.
\newblock {First Observation of the Oscillation Effect in the Neutrino-4
  Experiment on the Search for the Sterile Neutrino}.
\newblock {\em Pisma Zh. Eksp. Teor. Fiz.}, 109(4):209--218, 2019.

\bibitem{JUNO:2015zny}
Fengpeng An et~al.
\newblock {Neutrino Physics with JUNO}.
\newblock {\em J. Phys. G}, 43(3):030401, 2016.

\bibitem{Berryman_2021}
Jeffrey~M. Berryman and Patrick Huber.
\newblock {Sterile Neutrinos and the Global Reactor Antineutrino Dataset}.
\newblock {\em JHEP}, 01:167, 2021.

\bibitem{estienne2019updated}
M.~Estienne et~al.
\newblock {Updated Summation Model: An Improved Agreement with the Daya Bay
  Antineutrino Fluxes}.
\newblock {\em Phys. Rev. Lett.}, 123(2):022502, 2019.

\bibitem{patrick_huber_2022_6683772}
Patrick Huber.
\newblock {Neutrino science and nuclear security}.
\newblock {https://doi.org/10.5281/zenodo.6683772}, June 2022.
\newblock Accessed: 2022-06-22.

\bibitem{nufluxHuber:2004}
Patrick Huber and Thomas Schwetz.
\newblock Precision spectroscopy with reactor antineutrinos.
\newblock {\em Phys. Rev. D}, 70:053011, Sep 2004.

\bibitem{nuflux:1981}
K.~Schreckenbach et~al.
\newblock Absolute measurement of the beta spectrum from 235u fission as a
  basis for reactor antineutrino experiments.
\newblock {\em Phys. Lett. B}, 99(3):251--256, 1981.

\bibitem{nuflux:1985}
K.~Schreckenbach et~al.
\newblock Determination of the antineutrino spectrum from 235u thermal neutron
  fission products up to 9.5 mev.
\newblock {\em Phys. Lett. B}, 160:325--330, 1985.

\bibitem{nuflux:1982}
F.~von Feilitzsch et~al.
\newblock Experimental beta-spectra from 239pu and 235u thermal neutron fission
  products and their correlated antineutrino spectra.
\newblock {\em Phys. Lett. B}, 118:162--166, 1982.

\bibitem{nuflux:1989}
A.A. Hahn et~al.
\newblock Antineutrino spectra from 241pu and 239pu thermal neutron fission
  products.
\newblock {\em Phys. Lett. B}, 218:365--368, 1989.

\bibitem{nufluxHaag2014}
N.~Haag et~al.
\newblock {Experimental Determination of the Antineutrino Spectrum of the
  Fission Products of $^{238}$U}.
\newblock {\em Phys. Rev. Lett.}, 112(12):122501, 2014.

\bibitem{Mueller:2011}
Th.~A. Mueller et~al.
\newblock Improved predictions of reactor antineutrino spectra.
\newblock {\em Phys. Rev. C}, 83(5), May 2011.

\bibitem{Huber:2011}
Patrick Huber.
\newblock On the determination of anti-neutrino spectra from nuclear reactors.
\newblock {\em Phys. Rev. C}, 84:024617, Aug 2011.

\bibitem{Hayen:2019eop}
L.~Hayen, J.~Kostensalo, N.~Severijns, and J.~Suhonen.
\newblock {First-forbidden transitions in the reactor anomaly}.
\newblock {\em Phys. Rev. C}, 100(5):054323, 2019.

\bibitem{Vogel1999}
P.~Vogel and J.~F. Beacom.
\newblock Angular distribution of neutron inverse beta decay, $\bar{\nu}_e + p
  \rightarrow e^+ + n$.
\newblock {\em Phys. Rev. D}, 60(5), Jul 1999.

\bibitem{dalat2014}
Nhi~Dien Nguyen et~al.
\newblock Results of operation and utilization of the dalat nuclear research
  reactor.
\newblock {\em Nuclear Science and Technology}, 4(1):1--9, Mar. 2014.

\bibitem{futureCNR:2020}
Nhi-Dien Nguyen, Kien-Cuong Nguyen, Ton-Nghiem Huynh, Doan-Hai-Dang Vo, and
  Hoai-Nam Tran.
\newblock Conceptual design of a 10 mw multipurpose research reactor using
  vvr-kn fuel.
\newblock {\em Science and Technology of Nuclear Installations}, 2020:7972827,
  Aug 2020.

\bibitem{vnnupower2022}
Nuclear power in vietnam.
\newblock \url{https://world-nuclear.org}.
\newblock Accessed: 2022-06-22.

\bibitem{Huber:2002mx}
Patrick Huber, Manfred Lindner, and Walter Winter.
\newblock {Superbeams versus neutrino factories}.
\newblock {\em Nucl. Phys.}, B645:3--48, 2002.

\bibitem{Thanh:2021inp}
N.~H.~Duy Thanh et~al.
\newblock {Multi-pixel Photon Counter for operating the tabletop cosmic-ray
  detector under loosely controlled conditions}.
\newblock {\em arXiv physics.ins-det 2106.08603}, 6 2021.

\end{thebibliography}

\end{document}